\documentclass[aps,pra,superscriptaddress,showpacs,twocolumn,10pt]{revtex4-1}
\usepackage{bm}
\usepackage{amsmath}
\usepackage{textcomp}
\allowdisplaybreaks
\usepackage{longtable}
\usepackage{dcolumn}
\usepackage{nicefrac}
\newcolumntype{.}{D{x}{}{-1}}

\newcolumntype{w}[1]{D{.}{.}{#1}}
\newcommand{\balpha}{\bm{\alpha}}
\newcommand{\bgamma}{\bm{\gamma}}
\newcommand{\bsigma}{\bm{\sigma}}
\newcommand{\bfx}{\bm{x}}
\newcommand{\bfz}{\bm{z}}
\newcommand{\bfp}{\bm{p}}

\newcommand{\bfq}{\bm{q}}
\newcommand{\bfr}{\bm{r}}

\newcommand{\bfl}{\bm{l}}
\newcommand{\Za}{{Z\alpha}}

\newcommand{\bLambda}{\bm{\Lambda}}

\usepackage{graphicx}
\usepackage{xcolor}

\newcommand{\vare}{\varepsilon}
\newcommand{\lbr}{\langle}
\newcommand{\rbr}{\rangle}
\newcommand{\intinf}{\int^{\infty}_{-\infty}}

\newcommand{\hp}{\hat{\bfp}}
\newcommand{\hx}{\hat{\bfx}}
\newcommand{\hq}{\hat{\bfq}}

\newcommand{\cross}[1]{#1\!\!\!/}

\begin{document}

\title{Self-energy screening effects in the $\bm{g}$ factor of Li-like ions}

\author{V. A. Yerokhin}
\affiliation{Center for Advanced Studies,
        Peter the Great St.~Petersburg Polytechnic University, Polytekhnicheskaya 29,
        St.~Petersburg 195251, Russia}
\affiliation{Max~Planck~Institute for Nuclear Physics, Saupfercheckweg~1, D-69117 Heidelberg,
Germany}

\author{K. Pachucki}
\affiliation{Faculty of Physics, University of Warsaw,
             Pasteura 5, 02-093 Warsaw, Poland}

\author{M. Puchalski}
\affiliation{Faculty of Physics, University of Warsaw,
             Pasteura 5, 02-093 Warsaw, Poland}
\affiliation{Faculty of Chemistry, Adam Mickiewicz University,
             Umultowska 89b, 61-614 Pozna{\'n}, Poland}

\author{C. H. Keitel}
\affiliation{Max~Planck~Institute for Nuclear Physics, Saupfercheckweg~1, D-69117 Heidelberg,
Germany}

\author{Z. Harman}
\affiliation{Max~Planck~Institute for Nuclear Physics, Saupfercheckweg~1, D-69117 Heidelberg,
Germany}

\begin{abstract}
We report an investigation of the self-energy screening effects for the $g$ factor of the
ground state of Li-like ions. The leading screening contribution of the relative order $1/Z$ is
calculated to all orders in the binding nuclear strength parameter $Z\alpha$ (where $Z$ is the
nuclear charge number and $\alpha$ is the fine-structure constant). We also extend the known
results for the $\Za$ expansion of the QED screening correction by deriving the leading
logarithmic contribution of order $\alpha^5\ln\alpha$ and obtaining approximate results for the
$\alpha^5$ and $\alpha^6$ contributions. The comparison of the two approaches yields a
stringent check of consistency of the two calculations and allows us to obtain improved
estimations of the higher-order screening effects.
\end{abstract}


\maketitle

\section{Introduction}

Measurements of the bound-electron $g$ factor in light H-like ions have recently reached the
fractional accuracy of few parts in $10^{-11}$ \cite{sturm:11,sturm:13:Si,sturm:14}. In a
combination with advanced theoretical calculations, these measurements provided the most accurate
determination of the electron mass as well as one of the best tests of the bound-state quantum
electrodynamic (QED) theory. Extensions of these tests towards heavier H-like ions are
anticipated in the future \cite{vogel:19}. The main obstacle for such extensions is presently on
the theory side, caused by the insufficiently known two-loop QED effects
\cite{pachucki:05:gfact,czarnecki:18,sikora:20}.

Accurate experiments were performed also on the $g$ factors of Li-like ions
\cite{wagner:13,koehler:16,glazov:19}. They provided sensitive tests of the QED theory of the
electron-correlation and relativistic nuclear recoil effects, probing QED beyond the
external-field approximation. Recently, the experiments were extended further to B-like ions
\cite{arapoglou:19}, providing the first $g$-factor measurement for the non-zero orbital angular
momentum states. In future, a combination of the $g$-factor measurements in different charge
states of the same element has a potential to provide an independent determination of the
fine-structure constant $\alpha$ \cite{shabaev:06:prl,yerokhin:16:gfact:prl}.

In order to match the experimental precision, theoretical investigations of atomic $g$ factors
should be performed to all orders in the nuclear binding strength constant $\Za$ (where $Z$ is
the nuclear charge number and $\alpha$ is the fine-structure constant). Such calculations are
often very demanding and require taking into consideration numerous effects
\cite{shabaev:02:li,glazov:04:pra}. A number of highly sophisticated calculations were performed
during the past decade, most notably, the calculations of the self-energy and vacuum-polarization
screening corrections \cite{volotka:09,glazov:10}, the two-photon exchange correction
\cite{volotka:14}, and the nuclear recoil effect \cite{shabaev:17:prl}. Despite the achieved
progress, further investigations are needed in order to match the experimental precision for
light Li-like ions.

In calculations performed to all orders in $\Za$, the electron-electron interaction is accounted
for by perturbation theory, with the expansion parameter $1/Z$. The leading term of this
expansion $\propto\!1/Z^0$ corresponds to the hydrogenic approximation, i.e., the approximation
of non-interacting electrons. The higher-order terms $\propto\!1/Z^{1}$, $1/Z^{2}$, etc. are
induced by the electron-electron interaction. The modification of the hydrogenic corrections by
the electron-electron interaction is often referred to as the screening effect. In the present
work we investigate the effect of the screening of the QED corrections, which presently induces
one of the largest uncertainties in the theoretical predictions of $g$ factors of light Li-like
ions \cite{yerokhin:17:gfact,glazov:19}.

First calculations of the QED screening effect \cite{yan:01:prl,yan:02:jpb} included only the
leading term of the $Z\alpha$ expansion and were applicable just for the lightest ions. The
forthcoming investigations \cite{glazov:04:pra,glazov:06:pla,cakir:20} approximately included
contributions of higher orders in $Z\alpha$, but the accuracy of these approximation was rather
low, leading to errors of the screening effects $\sim$10\% for medium-$Z$ ions.

The first full-scale QED calculation of the self-energy and vacuum-polarization screening effects
was accomplished in Refs.~\cite{volotka:09,glazov:10}. These calculations accounted for the
leading screening corrections of the relative order $\propto\!1/Z$ rigorously and the
higher-order effects $\propto\!1/Z^{2+}$ approximately. Still, the numerical uncertainty of these
calculations $\sim$1-2\% was not sufficient for matching the experimental precision in the
low-$Z$ region. Moreover, the results were reported only for four ions, thus not allowing to
perform a consistency check between the all-order and the $\Za$-expansion calculations.

The main goal of the present work is to perform an independent calculation of the self-energy
screening correction for the $g$ factor of the ground state of Li-like ions. We aim to
cross-check the previously publishes results, to improve the numerical accuracy, and to perform a
detailed analysis of consistency of the all-order numerical approach against the $\Za$-expansion
calculations. To achieve this, we extend the existing $\Za$-expansion results by deriving the
leading logarithmic contribution of order $\alpha^5\ln\alpha$ and obtaining approximate results
for the $\alpha^5$ and $\alpha^6$ contributions. Combining the two methods, we obtain improved
estimations for the higher-order screening effects $\propto\!1/Z^{2+}$ and increase the accuracy
of the theoretical description of the QED screening effects in light Li-like ions.

The relativistic units ($\hbar=c=m=1$) and the Heaviside charge units ($ \alpha = e^2/4\pi$,
$e<0$) will be used throughout this paper.

\section{$\bm g$ factor}
\label{sec:gfact}

The linear Zeeman shift of the energy of an atomic state $v$  can be written as
\begin{align}
\delta E_v = g\, \mu_0 \, B\, \mu_v\, \,,
\end{align}
where $\mu_0 = |e|/(2m)$ is the Bohr magneton, $B = |\bm{B}|$ is the external magnetic field, $g$
is the $g$ factor of the atomic state, and $\mu_v$ is the angular-momentum projection on the
direction of the magnetic field. In the present work we assume that the nucleus has zero spin, so
that all interaction with the magnetic field comes from the electrons.

The relativistic interaction of an electron with the magnetic field is represented by an operator
\begin{align}
V_{\rm mag}(\bfr) = -e\balpha \cdot {\bm{A}}(\bfr) = \frac{|e|}{2}\,B\,(\bfr \times \balpha)_z\,,
\end{align}
where $\bm{A}(\bfr) = (\bm{B}\times\bfr)/2$ is the vector potential and we choose the $z$ axis to
be directed along $\bm{B}$. Expressing the energy shift caused by $V_{\rm magn}$ in terms of the
$g$ factor and fixing the angular-momentum projection of the atomic state as $\mu_v = 1/2$, we
introduce the effective operator responsible for the $g$ factor as
\begin{align}
V_g = 2\, (\bfr \times\balpha)_z\,.
\end{align}

The matrix element of the operator $V_g$ between two Dirac wave functions is evaluated as
\begin{align} \label{00}
\lbr n_1 | V_g| n_2 \rbr = (-1)^{j_1-\mu_1}\,C^{\,1\,0}_{j_2\,\mu_2,j_1\,-\mu_1}\, P(n_1n_2)\,,
\end{align}
where $j$ and $\mu$ are the total angular momentum and its projection, respectively,
$C_{j_1\mu_1, j_2\mu_2}^{jm}$ is the Clebsch-Gordan coefficient, and the radial integral $P$ is
given by
\begin{align}
P(n_1n_2) = &\ 2\, \frac{-\kappa_1-\kappa_2}{\sqrt{3}}\, C_1(-\kappa_2,\kappa_1)
 \nonumber \\ & \times
 \int_0^{\infty}dr\,r^3\,\big[ g_{n_1}(r)\,f_{n_2}(r)+f_{n_1}(r)\,g_{n_2}(r)\big]\,.
\end{align}
Here, $\kappa$ is the relativistic angular-momentum quantum number, $C_L(\kappa_a,\kappa_b)$ is
the reduced matrix element of the normalized spherical harmonics (see, e.g., Eq.~(C10) of
Ref.~\cite{yerokhin:99:pra}), and $g(r)$ and $f(r)$ are the upper and the lower radial components
of the Dirac wave function defined as in Ref.~\cite{yerokhin:99:pra}.

For the point-like nucleus, the diagonal matrix element of $V_g$ with hydrogenic Dirac wave
functions can be evaluated analytically as
\begin{align}
\lbr v | V_g| v \rbr = \frac{\kappa_v}{2j_v(j_v+1)} \Big(2\kappa_v \frac{\vare_v}{m}-1\Big)\,,
\end{align}
where $\vare_v$ is the Dirac energy. In particular, for the case relevant for this work of $v$
being the $2s$ state,
\begin{align}
\lbr 2s | V_g| 2s \rbr = \frac23 \Big( \sqrt{2\gamma+2}+1\Big)\,,
\end{align}
where $\gamma = \sqrt{1-(\Za)^2}$.

\section{General formulas}
\label{sec:general}

We now turn to the general formulas describing the self-energy screening correction to the $g$
factor of a Li-like ion. We will assume that the electronic configuration has the form of one
valence electron state (denoted by $v$) over a closed shell of core electron states (denoted by
$c$). The derivation of the formulas was first presented in Ref.~\cite{glazov:10} within the
formalism of the two-time Green function method \cite{shabaev:02:rep}. In the present work, we
will reformulate this problem in order to suit our calculational approach.

We start with introducing two operators which will be building blocks in the following formulas.
The first one is the electron-electron interaction operator $I(\omega)$, defined as
\begin{equation}\label{a1}
  I(\omega,\bfr_{1},\bfr_{2}) = e^2\, \alpha_{1}^{\mu} \alpha_{2}^{\nu}\, D_{\mu\nu}(\omega,\bfr_{12})\,,
\end{equation}
where $\alpha^{\mu} = (1,\balpha)$ are the Dirac matrices, $\bfr_{12} = \bfr_{1} - \bfr_{2}$, and
$D_{\mu\nu}(\omega,\bfr_{12})$ is the photon propagator. In the present work we use the Feynman
gauge, in which the photon propagator takes the simplest form,
\begin{equation}
  D_{\mu\nu}(\omega,\bfr_{12}) = g_{\mu\nu}\,
  \frac{e^{i\sqrt{\omega^2+i\epsilon}\,r_{12}}}{4\pi r_{12}}\,,
\end{equation}
where $r_{12} = |\bfr_{12}|$ and $\epsilon$ is a positive infinitesimal addition.

The one-loop self-energy (SE) operator $ \Sigma(\vare)$ is defined by its matrix elements with
the one-electron wave functions $|a\rbr$ and $|b\rbr$,
\begin{align}
\lbr a | \Sigma(\vare)| b\rbr = \frac{i}{2\pi}\intinf d\omega \sum_n
  \frac{\lbr an| I(\omega)| nb\rbr}{\vare-\omega - u\vare_n}\,,
\end{align}
where the sum over $n$ is carried out over the complete spectrum of the Dirac equation (implying
the summation over the discrete part of the spectrum and the integration over the continuum part
of the spectrum) and $u = 1 -i\epsilon$.

We will split the total self-energy screening correction into four parts as
\begin{align}\label{eq:sescr}
\Delta g_{\rm sescr} = \Delta g_{\rm po} + \Delta g_{\rm vr,Zee} + \Delta g_{\rm vr,scr} + \Delta g_{\rm dvr}\,,
\end{align}
with the individual contributions defined in the remaining of this Section.

\begin{widetext}

\subsection{Perturbed-orbital SE contribution}

The perturbed-orbital SE contribution incorporates all terms that can be expressed as matrix
elements of the one-loop SE operator $\Sigma(\vare)$. It can be represented as a sum of two
parts,
\begin{align}\label{eq000}
\Delta g_{\rm po} =  \Delta g_{\rm po1} + \Delta g_{\rm po2}\,,
\end{align}
where the first part contains matrix elements of the SE operator with a perturbed wave functions
on one side, whereas the second term has perturbed wave functions on both sides. The first term
can be expressed as
\begin{align} \label{eq000b}
\Delta g_{\rm po1} = 2\,\lbr v | \Sigma(\vare_v) | \delta_{\rm po1}  v\rbr
+ 2\,\lbr c | \Sigma(\vare_c) | \delta_{\rm po1} c\rbr
\,,
\end{align}
where
\begin{align}
|\delta_{\rm po1}  v\rbr \equiv \delta  |vcvc\rbr - \delta  |vccv\rbr\,, \ \ \ \
|\delta_{\rm po1}  c\rbr \equiv \delta  |cvcv\rbr - \delta  |cvvc\rbr\,,
\end{align}
and
\begin{align} \label{eq001}
\delta  |abcd\rbr  = & \sum_{\mu_{\rm core}} \Bigg\{
\sum_{n_1n_2}{\!}^{'}
\Bigg[
 \frac{|n_1\rbr\lbr n_1 | V_g |n_2 \rbr \lbr n_2b|I(\Delta_{db})|cd\rbr}
  {(\vare_a-\vare_{n_1})(\vare_a-\vare_{n_2})}
+
 \frac{|n_1\rbr  \lbr n_1b|I(\Delta_{db})|n_2d\rbr \lbr n_2 | V_g |c\rbr}
  {(\vare_a-\vare_{n_1})(\vare_c-\vare_{n_2})}
  \nonumber \\
& \ \ \  \ \ \ \  \ \ \ \ \ \ \ \ \ \
+ \frac{|n_1\rbr\lbr b | V_g |n_2 \rbr \lbr n_1n_2|I(\Delta_{db})|cd\rbr}
  {(\vare_a-\vare_{n_1})(\vare_b-\vare_{n_2})}
+
 \frac{|n_1\rbr \lbr n_1b|I(\Delta_{db})|cn_2\rbr \lbr n_2 | V_g |d \rbr }
  {(\vare_a-\vare_{n_1})(\vare_d-\vare_{n_2})}
\Bigg]
  \nonumber \\
%
%
& + \sum_{n}{\!}^{'}
 \Bigg[
 -\frac{|n\rbr \lbr n | V_g |a \rbr \lbr ab|I(\Delta_{db})|cd\rbr}{(\vare_a-\vare_{n})^2}
 -\frac{|n\rbr \lbr nb|I(\Delta_{db})|cd\rbr \lbr a | V_g |a \rbr}{(\vare_a-\vare_{n})^2}
 -\frac{|a\rbr \lbr a | V_g |n \rbr \, \lbr n b|I(\Delta_{db})|cd\rbr}{(\vare_a-\vare_{n})^2}
 \nonumber \\
&
\ \ \ \ \ \ \ \ \ \ \
+
\frac{|n\rbr \lbr n | V_g |a \rbr \lbr ab|I'(\Delta_{db})|cd\rbr }{\vare_a-\vare_{n}}
+
\frac{|n\rbr \lbr nb|I'(\Delta_{db})|cd\rbr \Big(\lbr d | V_g |d \rbr - \lbr b | V_g |b \rbr\Big)}{\vare_a-\vare_{n}}
 \nonumber \\
 & \ \ \ \ \ \ \ \ \ \ \
+
\frac{|a\rbr \lbr ab|I'(\Delta_{db})|nd\rbr \lbr n | V_g |c \rbr}{\vare_c-\vare_{n}}
+
\frac{|a\rbr \lbr ab|I'(\Delta_{db})|cn\rbr \lbr n | V_g |d \rbr}{\vare_d-\vare_{n}} \,
\Bigg]
  \nonumber \\
%
%
& +
\frac12\,|a\rbr \, \lbr ab|I''(\Delta_{db})|cd\rbr \, \Big(\lbr d | V_g |d \rbr - \lbr b | V_g |b \rbr\Big)
\Bigg\}\,,
\end{align}
where $a$, $b$, $c$, and $d$ are the one-electron states of the core or the valence electron.
Here and in what follows, $\Delta_{ab} = \vare_a-\vare_b$, the prime on the summation symbol
means that terms with vanishing denominator should be omitted from the summation, and each prime
in $I'(\omega)$ and $I''(\omega)$ denotes the derivative over the energy argument. The summation
over $\mu_{\rm core}$ runs over the angular-momentum projections of the core electron states,
$\mu_{\rm core} = \pm \nicefrac12$ for the $(1s)^2$ shell.

The second term in Eq.~(\ref{eq000}) is represented by
\begin{align}\label{eq0aa}
\Delta g_{\rm po2} = 2\,\sum_{\mu_c}\Big[
\big< \delta_{\rm Zee} v \big| \Sigma(\vare_v) \big| \delta_{\rm scr} v\big>
+
\big< \delta_{\rm Zee} c \big| \Sigma(\vare_c) \big| \delta_{\rm scr} c\big>
\Big]
\,,
\end{align}
where $\mu_c$ denotes the angular-momentum projection of the core electron state $c$, the
perturbed wave functions are defined by
\begin{align}
\big| \delta_{\rm Zee} a \big> = \sum_{n}{\!}^{'} \frac{|n\rbr \lbr n| V_g|a\rbr}{\vare_a-\vare_n}\,,
\ \ \ \
\big| \delta_{\rm scr} a \big> = \sum_{n}{\!}^{'} \frac{|n\rbr
 \big[ \lbr nb| I(0)|ab\rbr - \lbr nb| I(\Delta_{ab})|ba\rbr \big]}{\vare_a-\vare_n}\,,
\label{eq0ac}
\end{align}
and $(ab) = (vc)$ or $(cv)$.

In the notations of Ref.~\cite{glazov:10}, $\Delta g_{\rm po1}$ corresponds to the sum of the A,
E, and G terms, and $\Delta g_{\rm po2}$ corresponds to the B term. Formulas
(\ref{eq000})-(\ref{eq0ac}) were derived in Ref.~\cite{glazov:10} by the two-time Green's
function method \cite{shabaev:02:rep}. They can be also obtained by the standard
Rayleigh-Schr\"odinger perturbation theory as demonstrated in Appendix~\ref{sec:app1}.

\subsection{Perturbed Zeeman-vertex contribution}

The perturbed Zeeman-vertex contribution incorporates terms that can be expressed as non-diagonal
matrix elements of the Zeeman vertex operator plus the corresponding reducible part. It is given
by
\begin{align}\label{eq2a1}
\Delta g_{\rm vr, Zee} = &\ 2\,\sum_{\mu_c}\bigg\{
\big< v \big| \Big[\Lambda_{\rm Zee}(\vare_v)
     + V_{g,vv}\,  \Sigma'(\vare_v) \Big]\big|  \delta v\big>
+ \big< c \big| \Big[\Lambda_{\rm Zee}(\vare_c)
     + V_{g,cc}\,  \Sigma'(\vare_c) \Big]\big|  \delta c\big>
\bigg\}
\,,
\end{align}
where $V_{g, aa} \equiv \lbr a|V_{g}|a\rbr$, $\Sigma'(\vare)$ denotes the derivative of the
self-energy operator over the energy argument $\vare$,
 and the perturbed wave function is defined as
\begin{align}\label{eq2a1a}
\big| \delta a \big> &\ = \sum_{n}{\!}^{'} \frac{|n\rbr
 \big[ \lbr nb| I(0)|ab\rbr - \lbr nb| I(\Delta_{ab})|ba\rbr \big]}{\vare_a-\vare_n}
 -\frac12\, |a\rbr \lbr ab| I'(\Delta_{ab})|ba\rbr\,,
\end{align}
with $(ab) = (vc)$ or $(cv)$. The matrix element of the Zeeman vertex operator (with the
corresponding reducible part) is given by
\begin{align} \label{eq2a2}
\big< a \big| \Lambda_{\rm Zee}(\vare_a) + V_{g,aa}\,  \Sigma'(\vare_a)\big| \delta a\big>
= &\
\frac{i}{2\pi}
  \intinf d\omega \,
\sum_{n_1n_2}
\frac{\lbr an_2| I(\omega)|n_1\delta a\rbr\,\big[ \lbr n_1|V_g|n_2\rbr - \lbr n_1|n_2\rbr\, \lbr a|V_g|a\rbr\big]}
  {(\vare_a-\omega-u\,\vare_{n_1})(\vare_a-\omega-u\,\vare_{n_2})}
\,.
\end{align}

In the notations of Ref.~\cite{glazov:10}, $\Delta g_{\rm vr, Zee}$ corresponds to the sum of the
C1+H1 terms and a part of the H3 term.

\subsection{Perturbed screened-vertex contribution}

The perturbed screened-vertex contribution is a part that can be expressed in terms of
non-diagonal matrix elements of the screened (i.e., two-electron) vertex operator plus the
corresponding reducible part. We represent it as a sum of the vertex and the reducible parts,
\begin{align}\label{eq3a0}
\Delta g_{\rm vr,\, scr} = \Delta g_{\rm ver,\, scr} + \Delta g_{\rm red,\, scr}\,.
\end{align}
The vertex part is
\begin{align}\label{eq3a1}
\Delta g_{\rm ver,\, scr} = 2\,\sum_{PQ}(-1)^{P+Q}\,\sum_{\mu_{c}}
 &\
 \bigg[ \big< Pv\,Pc \big| \Lambda_{\rm scr} \big| \delta Qv\,Qc\big>
       + \big< Pv\,Pc \big| \Lambda_{\rm scr} \big| Qv\,\delta Qc\big>
\nonumber \\ &
       + \frac12\,
            \big< PvPc \big| \Lambda_{\rm scr.d} \big|  QvQc\big>
                  \,\Big( \lbr Qc|V_g|Qc\rbr - \lbr Pc|V_g|Pc\rbr \Big)
\bigg]
\,.
\end{align}
Here, $P$ and $Q$ are the permutation operators interchanging the valence and the core electrons,
$(PvPc) = (vc)$ or $(cv)$, $(QvQc) = (vc)$ or $(cv)$, $(\delta Qv\,Qc) = (\delta v\,c)$ or
$(\delta c\, v)$, $(Qv\,\delta Qc) = (v\,\delta c)$ or $( c\, \delta v)$, $(-1)^P$ and $(-1)^Q$
are the sign of the permutation $P$ and $Q$, respectively, $|\delta a\rbr \equiv |\delta_{\rm
Zee} a\rbr$ is the first-order perturbation of the wave function by the magnetic potential as
given in Eq.~(\ref{eq0ac}), and matrix elements of the two-electron vertex operator and its
derivative are defined by
\begin{align}
\lbr ab|{ \Lambda}_{{\rm scr}}|cd\rbr = &\ \frac{i}{2\pi}
  \intinf d\omega \,
\sum_{n_1n_2}
\frac{\lbr an_2| I(\omega)|n_1c\rbr\, \lbr n_1b|I(\Delta_{db})|n_2d\rbr}
  {(\vare_a-\omega-u\,\vare_{n_1})(\vare_c-\omega-u\,\vare_{n_2})}\,,
  \\
\lbr ab|{ \Lambda}_{{\rm scr.d}}|cd\rbr = &\ \frac{i}{2\pi}
  \intinf d\omega \,
\sum_{n_1n_2}
\frac{\lbr an_2| I(\omega)|n_1c\rbr\, \lbr n_1b|I'(\Delta_{db})|n_2d\rbr}
  {(\vare_a-\omega-u\,\vare_{n_1})(\vare_c-\omega-u\,\vare_{n_2})}\,.
\end{align}

The reducible part is defined as
\begin{align} \label{eq3a2}
\Delta g_{\rm red,\, scr} = 2\,\Big[ \lbr v|\Sigma'(\vare_v)|\widetilde{\delta v}\rbr
 + \lbr c|\Sigma'(\vare_c)|\widetilde{\delta c}\rbr
 \Big]
 \,,
\end{align}
where
\begin{align}\label{eq3a3}
|\widetilde{\delta a}\rbr = \sum_{\mu_{\rm core}}\bigg\{  &\
    |\delta a\rbr \, \big[\lbr ab|I(0)|ab\rbr -\lbr ab|I(\Delta_{ab})|ba\rbr\big]
    \nonumber \\ &
    + |a\rbr \, \Big[ \lbr \delta ab|I(0)|ab\rbr + \lbr a\delta b|I(0)|ab\rbr
     - \lbr \delta ab|I(\Delta_{ab})|ba\rbr - \lbr a\delta b|I(\Delta_{ab})|ba\rbr
    \nonumber \\ &
   \ \ \ \ \ \ \ \ \ \  -\frac12 \lbr ab|I'(\Delta_{ab})|ba\rbr \, \big(\lbr a|V_g|a\rbr - \lbr b|V_g|b\rbr\big)\Big] \bigg\}
   \,.
\end{align}
Here, $|\delta a\rbr \equiv |\delta_{\rm Zee} a\rbr$ and $(ab) = (vc)$ or $(cv)$.

In the notations of Ref.~\cite{glazov:10}, $\Delta g_{\rm ver,\, scr}$ corresponds to the sum of
the C2$+$H2$+$F terms, and $\Delta g_{\rm red,\, scr}$ corresponds to a part of the H3 term.

\subsection{Double-vertex contribution}

The double-vertex contribution is comprised of the matrix element of the double-vertex operator
plus the corresponding reducible parts, all of them containing the third power of ${\omega}$ in
the denominator. It is represented as
\begin{align} \label{eq4a0}
\Delta g_{\rm dvr} = \sum_{PQ}(-1)^{P+Q} \, \sum_{\mu_{c}}
 \big< PvPc\,\big|\, { \Lambda}_{\rm dvr}\,\big|\,QvQc\big>\,,
\end{align}
where the operator ${ \Lambda}_{\rm dvr}$ consists of four parts,
\begin{align} \label{eq4a1}
 { \Lambda}_{\rm vr,dbl} =  2\,{ \Lambda}_{\rm dver} +  2\,{ \Lambda}_{\rm d.scr} + { \Lambda}_{\rm d.Zee}
  + { \Lambda}_{\rm dd.se}\,.
\end{align}
The first term in the sum is the double-vertex operator, which is defined by its matrix element
as
\begin{align}\label{eq:verdbl}
2\,\lbr ab|\,{ \Lambda}_{\rm dver}\,|cd\rbr =  &\ 2\,\frac{i}{2\pi}\intinf d\omega \,
\sum_{n_1n_2n_3}
\frac{\lbr an_3| I(\omega)|n_1c\rbr\, \lbr n_1b|I(\Delta_{db})|n_2d\rbr\, \lbr n_2|V_{g}|n_3\rbr}
  {(\vare_a-\omega-u\,\vare_{n_1})(\vare_c-\omega-u\,\vare_{n_2})(\vare_c-\omega-u\,\vare_{n_3})}
\,.
\end{align}
The second term is the derivative of the screened-vertex operator, whose matrix element is
\begin{align}
2\,\lbr ab|\,{ \Lambda}_{\rm d.scr}\,|cd\rbr = &\ 2\,\lbr c|V_{g}|c\rbr\,\, \frac{i}{2\pi}
  \intinf d\omega \,
  \frac{\partial}{\partial\vare_c}
\sum_{n_1n_2}
\frac{\lbr an_2| I(\omega)|n_1c\rbr\, \lbr n_1b|I(\Delta_{db})|n_2d\rbr}
  {(\vare_a-\omega-u\,\vare_{n_1})(\vare_c-\omega-u\,\vare_{n_2})}
\,.
\end{align}
The third term is the derivative of the Zeeman-vertex operator, which is
\begin{align}
\lbr ab|\,{ \Lambda}_{\rm d.Zee}\,|cd\rbr = &\ \lbr ab|I(\Delta_{db})|cd\rbr\,\, \frac{i}{2\pi}
  \intinf d\omega \,
  \frac{\partial}{\partial\vare_a}
\sum_{n_1n_2}
\frac{\lbr an_2| I(\omega)|n_1a\rbr\, \lbr n_1|V_g|n_2\rbr}
  {(\vare_a-\omega-u\,\vare_{n_1})(\vare_a-\omega-u\,\vare_{n_2})}
\,.
\end{align}
The last term in Eq.~(\ref{eq4a1}) is the second derivative of the SE operator,
\begin{align}\label{eq4a10}
\lbr ab|\,{ \Lambda}_{\rm dd.se}\,|cd\rbr = &\ \lbr a|V_g|a\rbr\,\lbr ab|I(\Delta_{db})|cd\rbr\,\, \frac{i}{2\pi}\,
  \intinf d\omega \,
\frac{\partial^2}{\partial^2\vare_a}
\sum_{n}
\frac{\lbr an| I(\omega)|na\rbr}
  {\vare_a-\omega-u\,\vare_{n}}
\,.
\end{align}
In the notations of Ref.~\cite{glazov:10}, the four terms in the right-hand-side of
Eq.~(\ref{eq4a1}) correspond to the D, I2, I1, and I3 terms, respectively.

\end{widetext}

\section{Divergencies}
\label{sec:IR}

General formulas for the individual contributions presented in the previous Section contain
divergencies, both of the ultraviolet (UV) and infrared (IR) kind. The UV divergencies appear in
contributions containing the first and the second power of $\omega$ in the denominator(s) inside
the radiative photon loop. According to the standard procedure \cite{snyderman:91}, UV
divergencies are covariantly regularized by isolating one or two first terms of the expansion of
the bound-electron propagators in terms of the interaction with the binding nuclear field. These
terms are calculated in momentum space within the dimensional regularization, whereas the
remainder is calculated in coordinate space using the partial-wave expansion of the
bound-electron propagators. The UV divergencies are identified in terms of one-loop
renormalization constants and cancelled when all individual contributions are added together. The
cancellation of UV  divergencies was demonstrated in Ref.~\cite{glazov:10} and does not need to
be repeated here. In practical calculations, it is sufficient just to replace the free SE
operator and the free one-loop vertex operator by their renormalized expressions.

We now turn to the IR divergencies, which have not been discussed in detail in
Ref.~\cite{glazov:10}. These divergencies occur when the denominators of the electron propagators
inside the radiative photon loop vanish at $\omega \to 0$. As we will show below, the IR
divergencies originate from terms of the form
\begin{align} \label{eq:ir0}
J_{\beta} \equiv \frac{i}{2\pi}\intinf d\omega\, \frac{\lbr ab|I(\omega)|ab\rbr }{(-\omega+i0)^{\beta}}\,,
\end{align}
with $\beta \ge 2$. In the present investigation, we will encounter IR divergent terms with
$\beta = 2$ and $\beta = 3$. It should be noted that the term with $\beta = 1$ (appearing, e.g.,
in the one-loop SE matrix element) is IR safe. In order to show this, it is sufficient to rotate
the left half of the $\omega$ integration contour on the right half-axis, $(-\infty,0) \to
(\infty-i0,-i0)$, where the small addition $-i0$ indicates that this part lies on the lower bank
of the cut of the photon propagator. On the upper bank of the cut of the photon propagator,
$\sqrt{\omega^2} = \omega$, whereas on the lower bank, $\sqrt{\omega^2} = -\omega$. Therefore,
\begin{align}
J_{1} = \frac{i\alpha}{2\pi}\int_0^{\infty} d\omega\, \frac{\lbr ab|\alpha_{1\mu}\alpha_2^{\mu}\,
 \big( e^{i\omega x_{12}} - e^{-i\omega x_{12}}\big) |ab\rbr }{-\omega+i0}\,,
\end{align}
which is obviously converging at $\omega \to 0$.

In order to evaluate the IR divergent integrals $J_{2}$ and $J_3$, we regularize the divergencies
by introducing a finite photon mass $\mu$ in the photon propagator, evaluate the integral over
$\omega$ analytically, and separate out the $\mu$-dependent divergent terms, as described in
Ref.~\cite{yerokhin:20:green}. The results for the IR divergent integrals (omitting terms
vanishing in the limit $\mu\to 0$) are given by
\begin{align}\label{eq:j2}
J_2 = &\ \frac{\alpha}{\pi} \Big( \ln \frac{\mu}{2}+\gamma\Big)
  + \frac{\alpha}{\pi}\,\lbr ab|\alpha_{1\mu}\alpha_{2}^{\mu}\,\ln x_{12}|ab\rbr\,,
\end{align}
\begin{align}\label{eq:j3}
J_3
= &\
  \frac{\alpha}{4\,\mu}
-\frac{\alpha}{4}\,\lbr ab|\alpha_{1\mu}\alpha_{2}^{\mu}\, x_{12}|ab\rbr\,,
\end{align}
where $\gamma$ is Euler's constant.

We now demonstrate the cancellation of the IR divergencies in the sum (\ref{eq:sescr}). It is
convenient to express the IR-divergent parts of individual contributions in the form
\begin{align}
\Delta g_{i,{\rm IR}} = \sum_{PQ}(-1)^{P+Q} \sum_{\mu_{c}} \lbr PvPc| \Lambda_{i,\rm IR}|QvQc\rbr\,,
\end{align}
where $i$ runs over the contributions described in Sec.~\ref{sec:general}. The perturbed-orbital
SE contribution (\ref{eq000}) does not contain any IR divergences. In the perturbed Zeeman-vertex
contribution (\ref{eq2a1}), IR divergencies intrinsically present in the vertex and reducible
parts cancel each other, so that the total expression is finite and does not require a separate
treatment. The other contributions in Sec.~\ref{sec:general} contain IR divergencies, identified
as follows:
\begin{widetext}
\begin{align}
\label{eq:ir1}
\lbr ab| \Lambda_{\rm ver, scr, IR}|cd\rbr  =&\
 \frac{\alpha}{\pi}\, \Big( \ln \frac{\mu}{2}+\gamma\Big)\,
 2\, \bigg[
  \lbr ab|I(\Delta_{db})| c\,\delta_{\rm Zee} d\rbr
  + \frac12 \,
   \lbr ab|I'(\Delta_{db})| c\,d\rbr\,
   \big(\lbr d|V_g|d\rbr - \lbr b|V_g|b\rbr\big)
   \bigg]\,,
 \\
\lbr ab| \Lambda_{\rm ver, red, IR}|cd\rbr  =&\
 \frac{\alpha}{\pi}\, \Big( \ln \frac{\mu}{2}+\gamma\Big)\,  (-2)\,
 \nonumber \\ & \times
 \Big[
  \lbr \delta_{\rm Zee}a\,b|I(\Delta_{db})| c\, d\rbr
  +   \lbr a\,\delta_{\rm Zee}b|I(\Delta_{db})| c\, d\rbr
  + \frac12 \,
   \lbr ab|I'(\Delta_{db})| c\,d\rbr\,\big(\lbr d|V_g|d\rbr - \lbr b|V_g|b\rbr\big)
   \Big]\,,
 \\
\lbr ab| \Lambda_{\rm dver, IR}|cd\rbr =&\  \frac{\alpha}{4\mu}\, 2\,\lbr
ab|I(\Delta)|cd\rbr \,\lbr c |V_g|c\rbr
 + \frac{\alpha}{\pi}\, \Big( \ln \frac{\mu}{2}+\gamma\Big)\, 2\, \lbr ab|I(\Delta)|\delta_{\rm Zee} c\,d\rbr\,,
  \label{eq:dbl:ir}
\end{align}
\begin{align}
\lbr ab| \Lambda_{\rm d.scr, IR}|cd\rbr = &\ \frac{\alpha}{4\mu}\,(-2)\,\lbr ab|I(\Delta)|cd\rbr \,\lbr c |V_g|c\rbr
 \,,
 \\
\lbr ab| \Lambda_{\rm d.Zee, IR}|cd\rbr = &\ \frac{\alpha}{4\mu}\,(-2)\,\lbr ab|I(\Delta)|cd\rbr \,\lbr a |V_g|a\rbr
 \,,
  \\
\lbr ab| \Lambda_{\rm dd.se, IR}|cd\rbr = &\ \frac{\alpha}{4\mu}\,2\,\lbr ab|I(\Delta)|cd\rbr \,\lbr a |V_g|a\rbr
 \,.
\label{eq:ir2}
\end{align}
It can be easily seen that the sum of all IR contributions (\ref{eq:ir1}) - (\ref{eq:ir2})
vanishes. In actual calculations, the IR-divergent contributions were isolated by introducing
point-by-point subtractions in the integrand and then evaluated analytically according to
Eqs.~(\ref{eq:j2}) and (\ref{eq:j3}). Specifically, the matrix element of the double-vertex
operator (\ref{eq:verdbl}) is represented as (without the IR part accounted for by
Eq.~(\ref{eq:dbl:ir}))
\begin{align} \label{eq:ir5}
2\,\lbr ab|{ \Lambda}_{\rm dver}|cd\rbr =  &\ 2\,\frac{i}{2\pi}\intinf d\omega \, \Bigg\{
\sum_{n_1n_2n_3}
\frac{\lbr an_3| I(\omega)|n_1c\rbr\, \lbr n_1b|I(\Delta_{db})|n_2d\rbr\, \lbr n_2|V_{g}|n_3\rbr}
  {(\vare_a-\omega-u\,\vare_{n_1})(\vare_c-\omega-u\,\vare_{n_2})(\vare_c-\omega-u\,\vare_{n_3})}
\nonumber \\
& \ \ \ \ \ \ \ \ \ \ \ \ \ \ \ \
 - \sum_{\mu_{a'}\mu_{c'}\mu_{c''}}
\frac{\lbr ac'| I(\omega)|a'c\rbr\, \lbr a'b|I(\Delta_{db})|c''d\rbr\, \lbr c''|V_{g}|c'\rbr}
  {(-\omega+i0)^3}
\nonumber \\
& \ \ \ \ \ \ \ \ \ \ \ \ \ \ \ \
 -
\sum_{\mu_{a'}\mu_{c'},n_2\ne c} \frac{\lbr ac'| I(\omega)|a'c\rbr\, \lbr a'b|I(\Delta_{db})|n_2d\rbr\, \lbr n_2|V_{g}|c'\rbr}
  {(-\omega+i0)^2(\vare_c-\vare_{n_2})}
  \Bigg\}
\nonumber \\
&
- \frac{\alpha}{2}\,
\sum_{\mu_{a'}\mu_{c'}\mu_{c''}}
\lbr ac'| \alpha_{1\mu}\alpha_2^{\mu}\,x_{12}|a'c\rbr\, \lbr a'b|I(\Delta_{db})|c''d\rbr\, \lbr c''|V_{g}|c'\rbr
\nonumber \\
&
+ \frac{2\,\alpha}{\pi}\,
\sum_{\mu_{a'}\mu_{c'},n_2\ne c} \frac{\lbr ac'| \alpha_{1\mu}\alpha_2^{\mu}\,\ln x_{12}|a'c\rbr\, \lbr a'b|I(\Delta_{db})|n_2d\rbr\, \lbr n_2|V_{g}|c'\rbr}
  {(\vare_c-\vare_{n_2})}
\,,
\end{align}
where $a'$ and $c'$ ($c''$) denote the electron states that differ from $a$ and $c$ only by the
angular-momentum projection, $\mu_{a'}$ and $\mu_{c'}$ ($\mu_{c''}$), correspondingly.

\end{widetext}

\section{Computation of individual contributions}

\subsection{Perturbed-orbital SE contribution}

The calculation of the perturbed-orbital SE contributions, given by Eqs.~(\ref{eq000b}) and
(\ref{eq0aa}), is naturally reduced to a computation of non-diagonal matrix elements of the SE
operator. In the present work, we use the numerical approach developed in
Ref.~\cite{yerokhin:05:se}, which has an important advantage of a rapid convergence of the
partial-wave expansion. The perturbed wave functions in Eqs.~(\ref{eq000b}) and (\ref{eq0aa})
were calculated with help of the finite basis-set for the Dirac equation constructed with
$B$-splines \cite{johnson:88}. We do not use the dual kinetic balance (DKB) method
\cite{shabaev:04:DKB} in the present work, since our calculations are performed with the point
nuclear model, for which the DKB approach is not applicable.

The calculation of the perturbed-orbital SE contributions is simplified by the fact that the
matrix element of the SE operator is diagonal in the relativistic angular-momentum quantum number
$\kappa$ and the angular-momentum projection $\mu$ of the external wave functions,
\begin{align}
\lbr a | \Sigma(\vare) | b\rbr = \delta_{\kappa_a\kappa_b}\, \delta_{\mu_a\mu_b}\,\big(\ldots\big) \,,
\end{align}
where $(\ldots)$ does not depend on the angular-momentum projections. Therefore,
Eq.~(\ref{eq000b}) involves the perturbed wave functions of just one angular symmetry.
Eq.~(\ref{eq0aa}) contains a summation over several angular symmetries of the perturbed wave
functions (three in the general case but just two in our case). We also note that the perturbed
wave functions $| \delta_{\rm po} a \rbr$ and $| \delta_{\rm scr} a \rbr$ contain imaginary parts
which contribute when combined with the imaginary part of the SE operator.

For actual calculations of Eq.~(\ref{eq0aa}), it is convenient to move the summation over $\mu_c$
into the definition of one of the perturbed wave functions. For the first matrix element in
Eq.~(\ref{eq0aa}), this can be done immediately. In order to do this in the second matrix
element, we fix the angular-momentum projection of the $c$ state in the magnetic perturbed wave
function as $\mu_c = 1/2$,
\begin{align}
\big| \delta_{\rm Zee} c \big> &\ \to \sum_{n}{\!}^{'} \frac{|n\rbr \lbr n| V_g|c, \mu_c = \nicefrac{1}{2}\rbr}{\vare_c-\vare_n}\,,
\end{align}
and move the summation over $\mu_c$ into the definition of $| \delta_{\rm scr} c \rbr$, together
with the appropriate factor $s_{\mu_c}$ that carries the dependence of the magnetic matrix
element $\lbr n| V_g|c\rbr$ on $\mu_c$,
\begin{align}
\big| \delta_{\rm scr} c \big> &\ \to \sum_{\mu_c}s_{\mu_c}
  \sum_{n}{\!}^{'} \frac{|n\rbr
 \big[ \lbr nv| I(0)|cv\rbr - \lbr nv| I(\Delta_{cv})|vc\rbr \big]}{\vare_c-\vare_n}\,,
\label{eq1ab}
\end{align}
where
\begin{align}\label{eq:smuc}
s_{\mu_c} = (-1)^{\mu_c-1/2}\,C_{j_n\mu_c,j_c-\mu_c}^{10}
        \Big[C_{j_n\nicefrac{1}{2},j_c\,\nicefrac{-1}{2}}^{10}\Big]^{-1}\,.
\end{align}
In this way, we reduce the number of matrix elements of the SE operator to be computed by half.

In order to check our numerical procedure for computation of the perturbed wave functions in
Eqs.~(\ref{eq000b}) and (\ref{eq0aa}), we replace the self-energy operator by the
vacuum-polarization potential, $\Sigma(\vare) \to V_{\rm VP}$, thus reproducing the corresponding
vacuum-polarization screening corrections, calculated previously in
Refs.~\cite{glazov:10,cakir:20}.

\subsection{Perturbed Zeeman-vertex contribution}

The perturbed Zeeman-vertex contribution, defined by Eq.~(\ref{eq2a1}), can be considered as a
non-diagonal generalization of the Zeeman-vertex correction for the hydrogenlike atom,
specifically, $\Delta g_{\rm vr}$ given by Eq.~(12) of Ref.~\cite{yerokhin:04}. The only
difference of Eq.~(\ref{eq2a1}) as compared to the hydrogenic case is that the reference-state
wave function on the right-hand side of the matrix element is replaced by the perturbed wave
function $|\delta a\rbr$ given by Eq.~(\ref{eq2a1a}). It should be mentioned that the perturbed
wave function has contributions from several angular symmetries (three in the general case but
only two in our case of $\kappa_v = \kappa_c = -1$) and an imaginary part, which contributes to
the final result.

Similarly to the perturbed-orbital SE contribution, it is convenient to move the summation over
the angular-momentum projection of the core electron $\mu_c$ in Eq.~(\ref{eq2a1}) into the
definition of the perturbed wave function. For the first matrix element $\lbr v|\ldots|\delta
v\rbr$, it can be done immediately. For the second matrix element $\lbr c|\ldots|\delta c\rbr$,
some manipulations are needed. Specifically, we observe that the dependence of the vertex matrix
element on the angular-momentum projections of the external wave functions can be factorized out
as
\begin{align} \label{eq2a3}
\big< c \big| \Lambda_{\rm Zee}(\vare_c) \big| c'\big>
 = (-1)^{j_c-\mu_c}\,C_{j_{c'}\mu_{c'},\, j_c\,-\mu_c}^{\,1\,0}\, \big(\ldots\big) \,,
\end{align}
where $(\ldots)$ does not depend on the angular-momentum projections. Therefore, we can fix the
momentum projection $\mu_c = 1/2$ in the matrix element and redefine the perturbed wave function
as
\begin{align}
|{\delta} c \big> \to \sum_{\mu_c} s_{\mu_c} |{\delta} c \big>
\,,
\end{align}
where $s_{\mu_c}$ is given by Eq.~(\ref{eq:smuc}).

The numerical evaluation of the perturbed Zeeman-vertex contribution is similar to the
calculation of the diagonal matrix elements for the hydrogenic atoms, described in details in
Refs.~\cite{yerokhin:04,yerokhin:17:pra:segfact}. Specifically, the whole contribution is
separated into three parts,
\begin{align}
\Delta g_{\rm vr, Zee} = \Delta g_{\rm vr, Zee}^{(0)}
 + \Delta g_{\rm vr, Zee}^{(1)} + \Delta g_{\rm vr, Zee}^{(2+)}\,,
\end{align}
where the superscript indicates the number of interactions with the binding Coulomb potential in
the electron propagators. The first two terms in the right-hand-side of the above equation are
evaluated in momentum space, without any partial-wave expansion. Only the last term containing
two and more interactions with the Coulomb field is calculated in coordinate space. Thanks to the
separation of the one-potential term $\Delta g_{\rm vr, Zee}^{(1)}$, the partial-wave expansion
of the remainder $\Delta g_{\rm vr, Zee}^{(2+)}$ converges rapidly and can be calculated to high
accuracy. The contributions $\Delta g_{\rm vr, Zee}^{(0)}$ and $\Delta g_{\rm vr, Zee}^{(1)}$
need some generalization as compared to the diagonal case because of different angular symmetries
of the perturbed wave function; the corresponding formulas were already derived in our
calculation of the self-energy correction to the magnetic shielding
\cite{yerokhin:11:prl,yerokhin:12:shield}.

We note that as long as we calculate the vertex and the reducible part together and use an
appropriate $\omega$-integration contour (consisting of the low- and high-energy parts), the
integrand in Eq.~(\ref{eq2a2}) has a smooth small-$\omega$ behaviour. The would-be IR divergences
in the vertex and reducible parts are cancelled numerically at a given $\omega$ in this approach.
Alternatively, the contribution with $\vare_{n_1} = \vare_{n_2} = \vare_a$ can be separated out
and evaluated analytically with help of formulas from Sec.~\ref{sec:IR}. We checked the
equivalence of both methods in order to test the consistency of our numerical procedure.

We mention here some of the further cross-checks of the numerical procedure made in order to
eliminate possible errors: (i) by replacing $|\delta a\rbr \to |a\rbr$ in Eq.~(\ref{eq2a2}) we
reproduced known results for the vertex and reducible diagonal matrix elements for the $1s$ and
$2s$ hydrogenic states \cite{yerokhin:04}; (ii) by replacing $\Lambda_{\rm Zee}(\vare_a) +
V_{g,aa}\,  \Sigma'(\vare_a) \to V_g$ in Eq.~(\ref{eq2a2}) we reproduced known results for the
one-photon exchange correction to the $g$ factor, both in coordinate and momentum space.

\subsection{Perturbed screened-vertex contribution}

The perturbed screened-vertex contribution, represented by Eqs.~(\ref{eq3a0})-(\ref{eq3a3}), is
similar to the screened vertex and reducible corrections to the Lamb shift calculated in
Refs.~\cite{yerokhin:99:sescr,artemyev:05:pra,kozhedub:10}. The general scheme of evaluation
remains the same as for the Lamb shift. Specifically, the vertex contribution is separated into
the free $(0)$ and many-potential $(1+)$ parts,
\begin{align}
\Delta g_{\rm ver, scr} = \Delta g_{\rm ver, scr}^{(0)}
 + \Delta g_{\rm ver, scr}^{(1+)}\,.
\end{align}
The free vertex part contains only free electron propagators; it is renormalized and calculated
in momentum space. The many-potential part contains one and more interactions with the binding
Coulomb field; it is calculated in coordinate space using the partial-wave expansion of the
electron propagators. For the reducible contribution, we separate the zero-potential and
one-potential contributions,
\begin{align}
\Delta g_{\rm red, scr} = \Delta g_{\rm red, scr}^{(0)}
 + \Delta g_{\rm red, scr}^{(1)}
 + \Delta g_{\rm red, scr}^{(2+)}\,.
\end{align}
The zero- and one-potential contributions are calculated in momentum space. The separation of the
one-potential contribution improves the convergence of the partial-wave expansion in the
many-potential reducible contribution.

The computation of the many-potential part is very similar to that for the Lamb-shift case. We
introduce perturbed wave functions defined by Eqs.~(\ref{eq0ac}) and (\ref{eq3a3}) and calculate
them by using the finite basis-set method \cite{johnson:88}. Note that several different
symmetries of the perturbed wave functions contribute to the the final result (three in the
general case but just two in our case).

Contrary to the many-potential part, the evaluation of the free part turned out to be different
from our previous calculations for the Lamb shift \cite{yerokhin:99:sescr,artemyev:05:pra}. The
difference is that for the Lamb shift, analytical formulas for the basic angular integrals were
derived using averaging over the angular-momentum projections of the valence state. In the
$g$-factor calculations, the angular-momentum projection of the valence-electron state is fixed.
Moreover, we need to account for the case when the angular symmetry of the perturbed wave
function is different from the angular symmetry of the reference state. For this reason, we
developed a generalized procedure for performing angular integrations in the momentum space. The
evaluation of the free screened-vertex matrix elements is described in Appendix~\ref{app:verscr}.

Our numerical calculations of the many-potential vertex and reducible parts were performed using
the analytical representation of the Dirac-Coulomb Green function in terms of the Whittaker
functions. Integrations over the radial variables were carried out by the numerical approach
described in detail in the recent review \cite{yerokhin:20:green}. The partial-wave expansion was
extended up to $|\kappa| = 30$; the remaining tail of the expansion was estimated by a polynomial
fitting in $1/|\kappa|$.

We mention here several cross-checks of the computational procedure made in order to eliminate
possible errors: (i) we checked that in the diagonal case, our calculations reproduce known
results for the Lamb shift \cite{yerokhin:99:sescr}; (ii) we also checked that by replacing the
radiatively corrected vertex by the plain vertex we reproduce known results for the one-photon
exchange correction to the $g$ factor, both in coordinate and momentum space. Specifically, as a
part of this test, we checked that the replacement $\Gamma_{R,\mu} \to \gamma_{\mu}$ in
Eq.~(\ref{eq9aa}) yields the matrix element of the electron-electron interaction operator
$I(\Delta)$ in the coordinate-momentum representation.

\subsection{Double-vertex contribution}

The computation of the double-vertex contribution is the most complicated part of the
calculation. We start our discussion with the last three terms in the right-hand side of
Eq.~(\ref{eq4a1}). These terms are induced by derivatives of the Zeeman-vertex, screened-vertex,
and self-energy operators. Each of these operators were already examined, so we need only to
evaluate the derivative. In actual calculations, we find it convenient to convert the derivative
$\partial/(\partial \vare_a)$ to $\partial/(\partial \omega)$ and to apply integration by parts,
moving the derivative to the photon propagator. Specifically, we use the following identities,
\begin{align}
  \intinf d\omega \,
\frac{\partial}{\partial\vare_a}
\sum_{n_1n_2\neq a}
\frac{\lbr an_2| I(\omega)|n_1a\rbr\, \lbr n_1|V_g|n_2\rbr}
  {(\vare_a-\omega-u\,\vare_{n_1})(\vare_a-\omega-u\,\vare_{n_2})}
\nonumber \\
  =
  \intinf d\omega \,
\sum_{n_1n_2\neq a}
\frac{\lbr an_2| I'(\omega)|n_1a\rbr\, \lbr n_1|V_g|n_2\rbr}
  {(\vare_a-\omega-u\,\vare_{n_1})(\vare_a-\omega-u\,\vare_{n_2})}\,,
\end{align}
and
\begin{align}
  \intinf d\omega \,
\frac{\partial^2}{\partial^2\vare_a}
 &\
 \sum_{n\neq a}
\frac{\lbr an| I(\omega)|na\rbr}
  {\vare_a-\omega-u\,\vare_{n}}
\nonumber \\
  & =
  \intinf d\omega \,
\frac{\partial}{\partial\vare_a} \sum_{n\neq a}
\frac{\lbr an| I'(\omega)|na\rbr}
  {\vare_a-\omega-u\,\vare_{n}}
\nonumber \\
  & =
  \intinf d\omega \,
\sum_{n\neq a}
\frac{\lbr an| I''(\omega)|na\rbr}
  {\vare_a-\omega-u\,\vare_{n}} \,.
\end{align}
The advantage of using the above formulas is that the derivative over the photon propagator can
be easily evaluated analytically, contrary to the derivative over the electron propagator. An
additional bonus from this transformation is that the behaviour of the transformed integrand is
smoother in the low-$\omega$ region.

All double-vertex contributions (\ref{eq:verdbl})-(\ref{eq4a10}) contain the third power of
$\omega$ in the denominator and thus are convergent in the ultraviolet region. Therefore, in
principle, one does not need to separate out the zero-potential contributions in them. However,
we find it advantageous to do so, since this subtraction improves the convergence of the
partial-wave expansion drastically in the low-$Z$ region.

Specifically, we separate out the contributions of the free-electron propagators from the last
three terms on the right-hand side of Eq.~(\ref{eq4a1}) and evaluate them in momentum space,
without any partial-wave expansion. Formulas for these zero-potential contributions are easily
obtained by differentiation of the corresponding expressions for the zero-potential
Zeeman-vertex, screened-vertex, and self-energy operators. A comparison of the numerical results
obtained in the high-$Z$ region with and without the subtraction was employed as a useful check
of our numerical procedure. In the low-$Z$ region, the convergence of the partial-wave expansion
becomes increasingly slower. Even with the subtraction, we had to extend the partial-wave
expansion up to $|\kappa_{\rm max}| = 60$, in order to reach the desired numerical accuracy. The
computation was carried out with help of the analytical representation of the Dirac-Coulomb Green
function \cite{yerokhin:20:green}.

The evaluation of the matrix elements of the double-vertex operator (\ref{eq:verdbl}) is the most
computationally intensive part. The straightforward approach is to compute it as it stands, after
separating IR divergencies according to Eq.~(\ref{eq:ir5}). However, this turns out to be
applicable only in the high-$Z$ region; for lower $Z$, the convergence of the partial-wave
expansion becomes excruciatingly slow. The standard way to accelerate the convergence would be to
separate out the zero-potential double-vertex contribution and to calculate it in momentum space,
which is a very cumbersome task. Fortunately, this is not really needed. It turns out that the
convergence of the partial-wave expansion can be greatly accelerated if one separates out only a
relatively simple part of the full zero-potential contribution; this is the approach used in the
present work.

Specifically, we introduce the subtraction term $\Delta g_{\rm dver}^{\rm(s)}$ which we obtain
from $\Delta g_{\rm dver}$ by applying the following prescriptions: ({\em i}) all bound-electron
propagators are replaced by the free electron propagators, ({\em ii}) only the direct
contribution is taken ($PvPc = vc$ and $QvQc = vc$), ({\em iii}) only the Coulomb part of the
electron-electron interaction is taken ($I(\Delta) \to \alpha/r_{12}$). We note that under these
restrictions the contribution in which the radiative loop is attached to the core-electron line
vanishes after the summation over the angular-momentum projections of the core electrons, so that
only the contribution with the radiative loop attached to the valence electron line survives.

The subtraction term defined in this way is represented in momentum space as (see Sec.~III of
Ref.~\cite{yerokhin:04})
\begin{eqnarray}\label{eq:subtr}
\Delta g_{\rm dver}^{\rm(s)} &&= 4im \int \frac{d\bfp_1\,
    d\bfp_2\,d\bfl}{(2\pi)^6}\,
     V_{\rm core}(\bfl)\,
\nonumber \\ && \times\,
   \overline{\psi}_v(\bfp_1)\,
   \left[\bLambda_{\rm dver}^{(0)}(p_1,l,p_2)
     \times {\bm{\nabla}}_{\bfl} \delta^3(\bfq) \right]_z
       \psi_v(\bfp_2) \,,
       \nonumber \\
\end{eqnarray}
where $\bfq = \bfp_1-\bfl-\bfp_2$, $V_{\rm core}$ is the Fourier-transformed potential of the
charge density of the core electrons,
\begin{eqnarray}
     V_{\rm core}(\bfp) = 2\, \frac{4\pi\alpha}{p^2}\,\int_0^{\infty} dz\,z^2\, j_0(pz) \big[ g_c^2(z) + f_c^2(z)\big]\,,
      \nonumber \\
\end{eqnarray}
and $\bLambda_{\rm dver}^{(0)}$ is the free double-vertex operator defined as
\begin{eqnarray}
\bLambda_{\rm dver}^{(0)}(p_1,l,p_2) &&= -4\pi i\alpha \int \frac{d^4 k}{(2\pi)^4}\,
  \gamma_{\sigma}\, \frac{\cross{p}_1-\cross{k}+m}{(p_1-k)^2-m^2}\,
    \gamma_0\,
\nonumber \\ && \times
    \frac{\cross{p}_1-\cross{k}-\cross{l}+m}{(p_1-k-l)^2-m^2}\,
      \bgamma\,  \frac{\cross{p}_2-\cross{k}+m}{(p_2-k)^2-m^2}\,
         \gamma^{\sigma}\,.
\nonumber \\
\end{eqnarray}
Here, $p_1$, $l$, and $p_2$ are 4-vectors with a fixed time component, $p_1 = (\vare_v,\bfp_1)$,
$l = (\vare_v,\bfl)$, $p_2 = (\vare_v,\bfp_2)$, $\cross{p} = p_{\mu}\gamma^{\mu}$, and
$\gamma^{\mu} = (\gamma^0,\bgamma)$ are the Dirac matrices.

We now observe that the subtraction term (\ref{eq:subtr}) can be obtained from the one-potential
vertex contribution given by Eq.~(43) of Ref.~\cite{yerokhin:04}, by replacing the nuclear
Coulomb potential $V_C(q) = -4\pi Z\alpha/q^2$ with the potential of the core charge density
$V_{\rm core}(q)$. We, therefore, just use the formulas derived in Sec.~IIIB of
Ref.~\cite{yerokhin:04} in order to compute the double-vertex subtraction contribution
(\ref{eq:subtr}) in momentum space.

The remainder represented by the difference $\Delta g_{\rm dver} - \Delta g_{\rm dver}^{\rm(s)}$
was calculated in coordinate space using the partial-wave expansion of the electron propagators.
The low--energy part of the remainder was computed using the finite basis-set method. The
high-energy part of the remainder was evaluated with the analytical representation of the Dirac
Green function. The radial integrations were computed by the numerical approach described in
detail in the review \cite{yerokhin:20:green}.

The number of partial waves included into the computation varied from $|\kappa_{\max}| = 50$ in
the high- and medium-$Z$ region to $|\kappa_{\max}| = 75$ for $Z \leq 10$. In order to
cross-check our numerical procedure, we calculated the low-energy part of the remainder in two
ways, using the analytical representation of the Green function and the finite-basis set
$B$-spline representation.

\section{NRQED expansion}
\label{sec:nrqed}

\begin{table}[t]
\caption{Numerical results for the Fermi (spin-factorized) matrix element of the $\delta$ function operator for
the ground state of Li-like ions, in a.u.
\label{tab:fermidelta}}
\begin{ruledtabular}
\begin{tabular}{lw{4.16}}
 $Z$ & \multicolumn{1}{c}{$\lbr \sum_a\delta^3(\bm{r}_a)\rbr_F$ }\\\hline\\[-5pt]
  3 &  0.231\,249\,661\,(10) \\
  4 &  0.994\,525\,337\,(20) \\
  5 &  2.504\,853\,26\,(15) \\
  6 &  4.998\,567\,10\,(10) \\
  7 &  8.713\,793\,95\,(8) \\
  8 & 13.889\,046\,88\,(8) \\
  9 & 20.762\,960\,7\,(2) \\
 10 & 29.574\,218\,2\,(2) \\
 11 & 40.561\,523\,7\,(2) \\
 12 & 53.963\,593\,0\,(8) \\
 13 & 70.019\,147\,6\,(2) \\
 14 & 88.966\,913\,0\,(2) \\
\end{tabular}
\end{ruledtabular}
\end{table}

The nonrelativistic quantum electrodynamics (NRQED)  \cite{caswell:86} is the most general
approach that allows a systematic derivation of the expansion of various atomic properties in
terms of the fine-structure constant $\alpha$. The expansion terms are represented as expectation
values of some effective operators on the many-electron nonrelativistic wave function of the
reference state. The NRQED approach is most successive in describing light few-body systems since
for them the Schr\"odinger equation can be solved to very high numerical accuracy by using the
so-called explicitly correlated basis sets \cite{yan:95,korobov:02}.

An important feature of the NRQED approach is that each term of the $\alpha$ expansion includes
the electron-electron interaction  (more exactly, the parameter $1/Z$) to all orders. In the
present work, we use this feature in order to access the contributions of order $1/Z^2$ and
higher, which are not accounted for by the method presented in the previous Sections, and in
order to cross-check our computation of the $1/Z$ screening correction without expansion in
$\Za$.

The NRQED expansion of the radiative QED contribution to the bound-electron $g$ factor of light
atoms has the form
\begin{align}\label{eq:Za0}
\delta g_{\rm QED} =  \alpha^3\,g^{(3)}
 + \alpha^5\,\ln\alpha\,\,g^{(5,{\rm log})}
+ \alpha^5\,\,g^{(5)} + \alpha^6\,\,g^{(6)} + \ldots
\end{align}
The leading contribution of order $\alpha^3$ comes from anomalous magnetic moment of electron.
The corresponding formulas were first derived by Hegstrom in 1970th \cite{hegstrom:73}. The
result is
\begin{align}\label{eq:Za1}
g^{(3)} = & \frac{1}{\pi}\,\sum_a \big< Q_a^{(3)}\big>_F\,,
\end{align}
where $a$ numerates the electrons in the atom and the operator $Q_a^{(3)}$ is, in atomic units,
\begin{eqnarray}
Q_a^{(3)} &=& \frac{1}{3}\,\biggl(-\frac{p_a^2}{2}+\frac{Z}{r_a}-\sum_{b\neq a}\frac{1}{r_{ab}}\biggr)\,,
\end{eqnarray}
and $\lbr .\rbr_F$ denotes the so-called Fermi (spin-factorized) radial matrix element, defined
for an arbitrary one-electron operator $H$ as
\begin{align}
 \big< \psi \big| \sum_a H_a\, \bsigma_a \big| \psi\big> =
  \sum_a \big< \phi \big| H_a \big| \phi\big>_F\,2\,\bm{S}\,,
\end{align}
where $\psi$ is the full wave function of the reference state ({\em i.e.}, the antisymmetrized
product of the spatial and the spin functions), $\phi$ is the spatial part of the wave function,
and $\bm{S} = \sum_a\bm{\sigma}_a$ is the spin operator. For more details we refer the reader to
Ref.~\cite{yerokhin:17:gfact}.

Numerical calculations of $g^{(3)}$ for Li and Li-like ions with $Z\le 14$ were carried out in
Refs.~\cite{yan:01:prl,yan:02:jpb,yerokhin:17:gfact}. The results can be represented in terms of
the $1/Z$ expansion as
\begin{align}
g^{(3)}
 = &\, \frac1{\pi} \bigg( \frac{1}{24}\,Z^2 - \frac{274}{2187}\, Z
 \nonumber \\ &
 + 0.070\,41 + 0.0017\,\frac1Z+\ldots\bigg)\,,
\end{align}
where the first two coefficients are known exactly \cite{yerokhin:17:gfact} and the others are
obtained by fitting the numerical data.

In the present work we calculate the next-order logarithmic contribution in Eq.~(\ref{eq:Za0}),
of order $\alpha^5\ln\alpha$. The result is
\begin{align}\label{eq:Za2}
g^{(5,{\rm log})} =
   -\frac{64}{9}\,Z \ln Z\, \Big< \sum_a \delta^3(\bm{r}_a)\Big>_F \,.
\end{align}
This formula is obtained as a straightforward generalization of the corresponding result for the
hydrogenic ions derived in Ref.~\cite{pachucki:04:prl}, by using the  substitution of the
hydrogenic electron density of the $2s$ state on the nucleus by the corresponding few-body
density,
\begin{align}\label{eq:sub}
\big< \delta^3(\bm{r})\big> \equiv \frac{Z^3}{8\pi} \to \Big< \sum_a\delta^3(\bm{r}_a)\Big>_F\,.
\end{align}
The Fermi matrix elements of the $\delta$ function are calculated in the present work and listed
in Table~\ref{tab:fermidelta}. The results can be also represented in the form of the $1/Z$
expansion as
\begin{align}\label{eq:delta}
\Big<\sum_a\delta^3(\bm{r}_a)\Big>_F
 = \frac{Z^3}{8\pi} \, \Big( 1 - 2.6557\,\frac{1}{Z} + 0.914\,\frac{1}{Z^2}+\ldots\Big)\,.
\end{align}

A calculation of the non-logarithmic contribution of order $\alpha^5$ for few-electron atoms is a
difficult problem since it involves a low-energy contribution, which is analogous to the Bethe
logarithm for the Lamb shift. In the present work, we devise an approximation for the $\alpha^5$
and $\alpha^6$ contributions in Eq.~(\ref{eq:Za0}), basing on known results for the hydrogenic
case. Specifically, we get
\begin{align}\label{eq:Za3}
g^{(5)} + \alpha\, g^{(6)} \approx
   Z\,\Big< \sum_a \delta^3(\bm{r}_a)\Big>_F
 \, \Big[
a_{40} + (\Za)\,a_{50}\Big]   \,.
\end{align}
This formula is obtained from the hydrogenic result of Ref.~\cite{pachucki:04:prl} by the
substitution (\ref{eq:sub}), with $a_{40}$ and $a_{50}$ being the hydrogenic coefficients to
orders $\alpha(\Za)^4$ and $\alpha(\Za)^5$ for the $2s$ state. The numerical results for the
hydrogenic coefficients are $a_{40,\rm se}(2s) = -10.707\,716$ \cite{pachucki:04:prl}, $a_{40,\rm
vp}(ns) = -16/15$ \cite{karshenboim:01:jetp}, $a_{50,\rm se}(ns) = 23.282\,005$
\cite{pachucki:17:fact}, and $a_{50,\rm vp}(ns) = 127\pi/216$
\cite{karshenboim:01:jetp,karshenboim:02:plb}, with subscripts ``se'' and ``vp'' labelling
contributions originating from the self-energy and vacuum-polarization, correspondingly.

The hydrogenic coefficient $a_{40}(ns)$ is weakly $n$-dependent (a $5\%$ difference between the
$1s$ and $2s$ states), so we expect the unknown screening contribution to it to be within
10-20\%. The next-order coefficient $a_{50}$ is $n$-independent, so the corresponding result for
few-electron atoms is exact.

\section{Results and discussion}

Numerical results of our all-order calculation of the self-energy screening correction to the $g$
factor of the ground $(1s)^22s$ state of Li-like ions are presented in Table~\ref{tab:sescr}. Our
calculation is performed for the point nucleus. We note significant numerical cancellations
between individual contributions, present throughout the whole $Z$ region. For $Z = 82$, our
result is in perfect agreement with that of Ref.~\cite{glazov:10} but significantly more
accurate.

\begin{figure}
\centerline{
\resizebox{\columnwidth}{!}{%
  \includegraphics{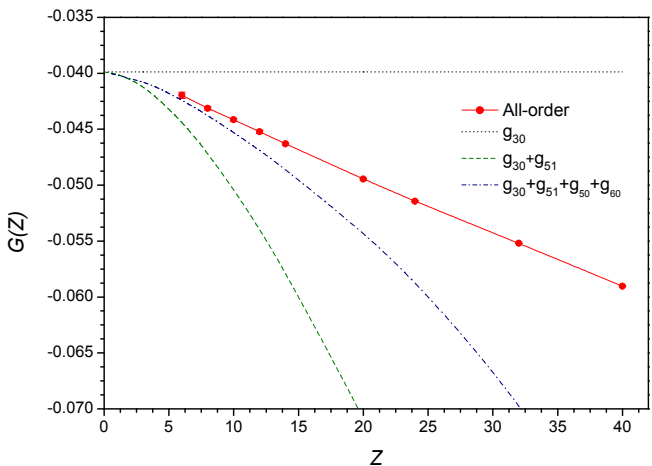}
} }
\vskip0.5cm
\caption{The self-energy screening correction to the $g$ factor of Li-like ions, in terms of the
scaled function $G(Z)$, defined by Eq.~(\ref{eq:Gfunc0}). The solid line and filled dots (red) represent
the numerical all-order (in $\Za$) results obtained in this work. The dotted line corresponds to
the contribution of the leading $\Za$-expansion term $g_{30}$. The dashed line (green) corresponds to the contribution
of two first terms of the $\Za$ expansion, $g_{30}$ and $g_{51}$. The dashed-dotted line (blue)
shows the contribution of all known terms in Eq.~(\ref{eq:Gfunc1}).
\label{fig:Gfunc}}
\end{figure}

In order to analyze our all-order numerical results for the self-energy screening correction, it
is convenient to separate out its leading $\alpha$ and $Z$ dependence, by introducing the
function $G(\Za)$ as follows
\begin{align}\label{eq:Gfunc0}
\Delta g_{\rm sescr} = \alpha^2 (\Za)\, G(\Za)\,.
\end{align}
The $Z\alpha$ expansion of the function $G$ follows from the results obtained in
Sec.~\ref{sec:nrqed},
\begin{align}\label{eq:Gfunc1}
G(\Za) = &\, g_{30} + (Z\alpha)^2
 \Big[ g_{51}\ln(Z\alpha)^{-2}
+ g_{50} + (\Za)\, g_{60} + \ldots\Big]\,,
\end{align}
where the leading coefficient $g_{30} = {-274}/(2187\pi)  = -0.039\,880\ldots$, $g_{51} =
-2.6557/(8\pi)\times(32/9)   = -0.3757$, $g_{50}\approx -2.6557/(8\pi)\times (-10.708)$, and
$g_{60} = -2.6557/(8\pi)\times 23.282$. Fig.~\ref{fig:Gfunc} shows the comparison of the
all-order numerical results for the scaled function $G(\Za)$ with the predictions based on the
$\Za$-expansion (\ref{eq:Gfunc1}). We conclude that our all-order results converge to the
prediction of the $\Za$-expansion as $Z \to 0$. We also observe that the inclusion of the
approximate higher-order contributions $g_{50}$ and $g_{60}$ significantly improves the agreement
between the all-order and $\Za$-expansion results. The deviation of the all-order results from
the $\Za$-expansion in the low-$Z$ region is consistent with an additional contribution to
$g_{50}$, $\delta g_{50} \approx 0.2$.

\begin{table*}
\caption{Self-energy screening correction to the
$g$ factor of the ground state of Li-like ions, for the point nucleus, in units of $10^{-6}$.
\label{tab:sescr}}
\begin{ruledtabular}
\begin{tabular}{l w{3.6}  w{3.6}  w{3.6}  w{3.6}  w{3.6}  w{3.6}  w{3.6}  w{3.6}  }
 \multicolumn{1}{c}{$Z$}   & \multicolumn{1}{c}{{\rm po}}
  & \multicolumn{1}{c}{{\rm vrzee}}  & \multicolumn{1}{c}{{\rm vrscr}}  & \multicolumn{1}{c}{{\rm d.Zee}}
  & \multicolumn{1}{c}{{\rm d.scr}}  & \multicolumn{1}{c}{{\rm dd.se}}  & \multicolumn{1}{c}{{\rm dver}}
  & \multicolumn{1}{c}{Total} \\
\hline\\[-5pt]
  6 &  1.3974\,(2)     &  -1.4021\,(2)    &  -6.2588\,(1)    &  17.1449\,(5)    &  13.3450\,(2)    &  -13.9006\,(1)   &  -10.4237\,(3)   &  -0.0979\,(7)   \\
  8 &  1.8551\,(2)     &  -1.8651\,(2)    &  -7.8317\,(1)    &  21.1762\,(2)    &  16.5302\,(2)    &  -17.2682\,(1)   &  -12.7307\,(2)   &  -0.1341\,(5)   \\
 10 &  2.3088\,(3)     &  -2.3259\,(2)    &  -9.2930\,(2)    &  24.8517\,(1)    &  19.4483\,(2)    &  -20.3670\,(1)   &  -14.7945\,(2)   &  -0.1715\,(5)   \\
 12 &  2.7588\,(3)     &  -2.7851\,(2)    &  -10.6671\,(1)   &  28.2509\,(1)    &  22.1588\,(1)    &  -23.2561\,(1)   &  -16.6711\,(5)   &  -0.2109\,(7)   \\
 14 &  3.2055\,(3)     &  -3.2431\,(2)    &  -11.9695\,(1)   &  31.4263\,(1)    &  24.7013\,(2)    &  -25.9750\,(1)   &  -18.3974\,(4)   &  -0.2520\,(6)   \\
 18 &  4.0917\,(2)     &  -4.1584\,(2)    &  -14.4012\,(1)   &  37.2484         &  29.3892\,(2)    &  -31.0088\,(1)   &  -21.4996\,(6)   &  -0.3388\,(7)  \\
 20 &  4.5325\,(2)     &  -4.6170\,(2)    &  -15.5459\,(2)   &  39.9462\,(1)    &  31.5731\,(2)    &  -33.3620\,(1)   &  -22.9113\,(4)   &  -0.3844\,(6)   \\
 24 &  5.4126\,(1)     &  -5.5391\,(2)    &  -17.7206\,(1)   &  45.0071\,(1)    &  35.6896\,(1)    &  -37.8097\,(2)   &  -25.5197\,(2)   &  -0.4797\,(4)   \\
 32 &  7.1871\,(1)     &  -7.4256\,(1)    &  -21.7175\,(3)   &  54.1335         &  43.1770\,(1)    &  -45.9303\,(2)   &  -30.1107\,(1)   &  -0.6864\,(4)   \\
 40 &  9.0175\,(1)     &  -9.4104\,(2)    &  -25.3893\,(3)   &  62.3769         &  50.0049\,(3)    &  -53.3559\,(2)   &  -34.1613        &  -0.9177\,(5)   \\
 54 &  12.4894\,(1)    &  -13.2895\,(1)   &  -31.3949\,(2)   &  75.6765         &  61.1161\,(3)    &  -65.4552\,(3)   &  -40.5563        &  -1.4139\,(5)   \\
 70 &  17.1547\,(1)    &  -18.7634\,(1)   &  -38.1666\,(3)   &  90.3570\,(2)    &  73.4067\,(4)    &  -78.8834\,(4)   &  -47.3792        &  -2.2741\,(7)   \\
 82 &  21.3885\,(2)    &  -24.0242\,(1)   &  -43.6095\,(4)   &  101.7123        &  82.8012\,(4)    &  -89.2703\,(5)   &  -52.3435        &  -3.3455\,(8)   \\
    &                &                &                &                &                &                &                &  -3.3^a\\
 92 &  25.5694\,(2)    &  -29.5452\,(2)   &  -48.6985\,(4)   &  111.8057        &  90.9525\,(3)    &  -98.4813\,(5)   &  -56.3789        &  -4.7763\,(8)   \\
\end{tabular}
$^a$ Glazov {\em et al.}, 2010~\cite{glazov:10}\,.
\end{ruledtabular}
\end{table*}

So far we addressed the self-energy screening correction of the relative order $1/Z$ as compared
to the leading, one-electron contribution. The complementary vacuum-polarization screening
correction was calculated previously in Ref.~\cite{glazov:10} and recently reproduced in
Ref.~\cite{cakir:20}.

In order to complete the calculation of the QED screening effects, we need also to estimate the
contribution of the higher-order screening $\propto 1/Z^{2+}$. For low-$Z$ ions, this can be done
immediately with help of the NRQED formulas presented in Sec.~\ref{sec:nrqed}. Such approach,
however, would result in large uncertainties for high-$Z$ ions. In order to avoid this, we
devised an estimate which is equivalent to the one obtained from NRQED in the low-$Z$ region but
applicable also for high-$Z$ ions. Specifically, we estimate the QED screening correction of
order $1/Z^2$ as
\begin{align}\label{eq:qedho1}
g^{(1/Z^2)}_{\rm QED} \approx \alpha^3\, \left[ 0.022412 + \frac{(\Za)^2}{8\pi}\,\frac{c_2}{c_1}\,H^{(1)}(\Za)\right]\,,
\end{align}
where the first term in the brackets comes from the $1/Z$ expansion of $\lbr Q_a\rbr$ in
Eq.~(\ref{eq:Za2}), $c_1 = -2.6557$ and $c_2 = 0.914$ are the coefficients of the $1/Z$ expansion
of $\lbr \sum_a\delta(\bfr_a)\rbr_F$ in Eq.~(\ref{eq:delta}), and $H^{(1)}$ is the $1/Z^1$
higher-order remainder extracted from the all-order results obtained in this work. Specifically,
we define $H^{(1)}$ by representing the all-order results for the self-energy and
vacuum-polarization screening of relative order $1/Z$ as
\begin{align}
g^{(1/Z)}_{\rm QED} = \alpha^2\,(\Za) \left[ -\frac{274}{2187\pi} + \frac{(\Za)^2}{8\pi}\,H^{(1)}(\Za)\right]\,.
\end{align}

We also devise an alternative estimation of the $1/Z^2$ QED screening contribution, based on the
{\em one-electron} QED results. Specifically, we introduce the $1/Z^0$ higher-order remainder
function $H^{(0)}$, by representing the all-order results for the {\rm one-electron} self-energy
and vacuum-polarization of the $2s$ state as
\begin{align}
g^{(1/Z^0)}_{\rm QED} = \frac{\alpha}{\pi}
 + \alpha\,(\Za)^2 \left[ \frac{1}{24\pi} + \frac{(\Za)^2}{8\pi}\,H^{(0)}(\Za)\right]\,.
\end{align}
After that, our second approximation for the $1/Z^2$ QED screening correction is obtained as
\begin{align}\label{eq:qedho2}
g^{(1/Z^2)}_{\rm QED} \approx \alpha^3\, \left[ 0.022412 + \frac{(\Za)^2}{8\pi}\,\frac{c_2}{c_0}\,H^{(0)}(\Za)\right]\,,
\end{align}
where $c_0 = 1$ is the first coefficient of the $1/Z$ expansion (\ref{eq:delta}).

The QED screening effects of order $1/Z^3$ and higher are relevant only for the lightest ions and
can be accounted for in the leading order of the $\alpha$ expansion,
\begin{align}\label{eq:qedho3}
g^{(1/Z^{3+})}_{\rm QED} = & \frac{\alpha^3}{\pi}\,\Big<\sum_aQ_a^{(3)}\Big>_F^{(3+)} \,,
\end{align}
where the matrix element contains all terms of its $1/Z$ expansion starting with $1/Z^3$,
\begin{align}
\Big<\sum_aQ_a^{(3)}\Big>_F^{(3+)} = &\, \Big<\sum_aQ_a^{(3)}\Big>_F
 \nonumber \\ &
- \Big(\frac1{24}Z^2
- \frac{274}{2187}\, Z
 + 0.070\,41 \Big)\,.
\end{align}

Table~\ref{tab:qedscr} summarizes our results for the individual QED screening contributions for
the ground state of Li-like ions. The column ``SE(pnt)'' lists results for the self-energy
screening correction of the relative order $1/Z$ for the point nuclear model, taken from
Table~\ref{tab:sescr}.

The column ``SE(fns)'' presents our results for the shifts of the point-nucleus correction due to
the finite nuclear size. It was calculated by taking the corresponding hydrogenic correction for
the $2s$ state, obtained as in Ref.~\cite{yerokhin:13:jpb}, and scaling it with help of
Eq.~(\ref{eq:sub}). We assumed an uncertainty of 25\% of this approximation. This correction is
relevant only for high-$Z$ ions; its error is negligible on the level of the total theoretical
uncertainty.

The column ``VP'' lists results for the vacuum-polarization screening correction of the relative
order $1/Z$, as calculated in Ref.~\cite{cakir:20}. The uncertainty of this part comes from
uncalculated higher-order contributions in the so-called magnetic-loop vacuum-polarization. It
was estimated by multiplying the results for the magnetic-loop correction in the free-loop
approximation (see Ref.~\cite{cakir:20} for details) by the factor of $2\,(\Za)^2$.

The column ``h.o.'' presents results for the QED screening correction of the relative order
$1/Z^{2+}$. They were obtained as a half-sum of the two approximations given by
Eqs.~(\ref{eq:qedho1}) and (\ref{eq:qedho2}) plus the higher-order screening contribution given
by Eq.~(\ref{eq:qedho3}). The quoted uncertainty was obtained as twice the difference between the
two approximations.

In Table~\ref{tab:qedscr} we compare our final theoretical values for the QED screening effect
with previous results obtained by Glazov, Volotka and co-authors \cite{volotka:14,glazov:19}. We
mention that our present approach is equivalent to theirs for the $1/Z$ part of the QED screening
effects but differs in estimating the higher-order $1/Z^{2+}$ screening contributions. In our
approach, we base our estimate on the NRQED results, whereas Refs.~\cite{volotka:14,glazov:19}
approximately accounted for the higher-order screening by using various screening potentials. We
observe the general consistency of the two calculations. This consistency is a rather strict
test, because of delicate numerical cancellations between the individual terms required to get
the final result. Still, the estimated error bars of the two calculations do not overlap in most
cases, the deviation being on the level of 2-3$\,\sigma$. This indicates that further work is
needed in order to fully cross-check the QED screening calculations.

\begin{table*}
\caption{QED screening corrections to the $g$ factor of the ground state of Li-like ions, in units of $10^{-6}$.
\label{tab:qedscr}}
\begin{ruledtabular}
\begin{tabular}{l w{3.6}  w{3.6}  w{3.6}  w{3.6}  w{3.6}  w{3.6}   w{3.6} }
 \multicolumn{1}{c}{$Z$}
  & \multicolumn{1}{c}{SE (pnt)}
  & \multicolumn{1}{c}{SE (fns)}
  & \multicolumn{1}{c}{VP}
  & \multicolumn{1}{c}{h.o.}
  & \multicolumn{1}{c}{Total}
  & \multicolumn{1}{c}{Ref.~\cite{glazov:19}}
  & \multicolumn{1}{c}{Ref.~\cite{volotka:14}}
  \\
\hline\\[-5pt]
  6 &  -0.097\,9\,(7)    &                    &  0.000\,5          &  0.009\,1\,(1)     &  -0.088\,3\,(7)   \\
  8 &  -0.134\,1\,(5)    &                    &  0.001\,2          &  0.009\,2\,(2)     &  -0.123\,7\,(5)   \\
 10 &  -0.171\,5\,(5)    &                    &  0.002\,3          &  0.009\,3\,(3)     &  -0.160\,0\,(6)   \\
 12 &  -0.210\,9\,(7)    &                    &  0.003\,8          &  0.009\,4\,(4)     &  -0.197\,6\,(8)   \\
 14 &  -0.252\,0\,(6)    &                    &  0.006\,0          &  0.009\,6\,(6)     &  -0.236\,4\,(8)    & -0.241\,5\,(21) & -0.236\,(5)   \\
 18 &  -0.338\,8\,(7)    &                    &  0.012\,5          &  0.009\,8\,(9)     &  -0.316\,5\,(11)  \\
 20 &  -0.384\,4\,(6)    &                    &  0.016\,9          &  0.010\,0\,(11)    &  -0.357\,5\,(12)   &                 & -0.370\,(7) \\
 24 &  -0.479\,7\,(4)    &                    &  0.028\,8          &  0.010\,2\,(15)    &  -0.440\,7\,(16)  \\
 32 &  -0.686\,4\,(4)    &  0.000\,1          &  0.066\,9\,(2)     &  0.010\,7\,(25)    &  -0.608\,7\,(25)  \\
 40 &  -0.917\,7\,(5)    &  0.000\,3\,(1)     &  0.130\,6\,(6)     &  0.011\,0\,(36)    &  -0.775\,8\,(37)  \\
 54 &  -1.413\,9\,(5)    &  0.002\,0\,(5)     &  0.335\,5\,(36)    &  0.011\,7\,(57)    &  -1.064\,7\,(68)  \\
 70 &  -2.274\,1\,(7)    &  0.013\,3\,(33)    &  0.827\,(16)       &  0.012\,6\,(84)    &  -1.422\,(19)    \\
 82 &  -3.345\,5\,(8)    &  0.050\,(12)       &  1.545\,(40)       &  0.014\,(10)       &  -1.737\,(43)     &                  & -1.91\,(4)    \\
 92 &  -4.776\,3\,(8)    &  0.158\,(40)       &  2.573\,(77)       &  0.014\,(11)       &  -2.030\,(87)     &                  & -2.18\,(6)    \\
\end{tabular}
\end{ruledtabular}
\end{table*}

In Table~\ref{tab:si} we present a compilation of all known binding corrections to the $g$-factor
of the ground state of Li-like silicon, $^{28}$Si$^{11+}$. As compared to the analogous table in
our previous investigation \cite{yerokhin:17:gfact}, the one-loop QED effects of the relative
order $1/Z$ and $1/Z^{2+}$ and the nuclear recoil contributions were updated. The screened QED
effects are calculated in the present work, whereas the nuclear recoil corrections of relative
orders $1/Z^0$, $1/Z^1$, and $1/Z^{2+}$ were calculated by Shabaev {\em et
al.}~\cite{shabaev:17:prl}. It should be mentioned that Shabaev {\em et al.} found a mistake in
the previous calculations of the recoil effect \cite{yan:01:prl,yan:02:jpb}, which resulted in a
small shift of theoretical values for this correction.

Our final theoretical value presented in Table~\ref{tab:si} is in agreement with the previous
theoretical results of Glazov {\em et al.}~\cite{glazov:19} and Volotka {\em et
al.}~\cite{volotka:14}. However, our result disagrees with the recent experimental value
\cite{glazov:19} by about five standard deviations. It is interesting that the theoretical
predictions, with time, are moving away from the experimental result, while steadily increasing
in estimated accuracy.

Considering the comparison of the present theoretical prediction for silicon with the latest
theoretical result of Glazov {\em et al.}~\cite{glazov:19}, we note two small deviations. One is
the difference of $+0.005\,(2)\times 10^{-6}$ in the QED screening effect, whereas the other is
the difference of $-0.003\,(3)\times 10^{-6}$ in the $1/Z^{3+}$ electron-correlation effect.
These two differences partially cancel each other, resulting in the total difference of the two
total theoretical values of $+0.002\,(4)\times 10^{-6}$, well within the quoted error bars.

Commenting on the disagreement with the experimental result for silicon, we note that not all
effects in the theoretical prediction has been confirmed by independent calculations so far.
Apart from the already mentioned small inconsistencies between the two calculations of the QED
screening correction and the $1/Z^{3+}$ electron-correlation effect, the two-photon exchange QED
effect has  so far been calculated by one group only \cite{volotka:14}. The other contributions
in the theoretical prediction seem to be under a better control. In particular, the nuclear
recoil effect was recently calculated rigorously within QED by Shabaev and co-workers
\cite{shabaev:17:prl}. They found a mistake in the earlier calculation by
Yan~\cite{yan:01:prl,yan:02:jpb}. After correcting this mistake, the extrapolation of Yan's
results (as in Ref.~\cite{yerokhin:17:gfact}) yields a result for silicon that agrees with that
of Shabaev and co-workers up to few parts in $10^{-10}$. So, this effect should not be
responsible for the deviation from the experimental value.

\section*{Summary}

We performed a calculation of the self-energy screening effects for the $g$ factor of the ground
state of Li-like ions. The contribution of the relative order $1/Z$ was calculated rigorously
within QED, to all orders in the binding nuclear strength parameter $Z\alpha$. The higher-order
screening contribution $\propto 1/Z^{2+}$ was calculated approximately, using results for the
coefficients of the $\alpha$ expansion obtained in this work in the framework of nonrelativistic
QED. In the result, we were able to improve the theoretical accuracy of the QED screening effects
in light Li-like ions and, as a consequence, the accuracy of the theoretical prediction of the
$g$ factor of Li-like silicon. The total theoretical result for Li-like silicon is in agreement
with previous theoretical calculations but differs by about five standard deviations from the
experimental result. We conclude that further theoretical and experimental work is needed in
order to investigate small deviations between different theoretical calculations and a much
larger discrepancy with the experimental result.

\begin{acknowledgments}
Work presented in this paper
was supported by the Russian Science Foundation (Grant No. 20-62-46006).
Valuable discussions with A.~Volotka and D.~Glazov are gratefully acknowledged.
\end{acknowledgments}

%
%
\begin{table}[t]
\caption{Binding corrections to the $g$-factor of the ground state of $^{28}$Si$^{ 11+}$.
The sum of all binding contributions is the difference of the atomic $g$ factor and the
free-electron $g$ factor, $g_e = 2.002\,319\,304\,361\,5\,(6)$
\cite{hanneke:08}. Nuclear parameters used in the calculation are: $M/m = 50\,984.832\,73$ and $R =
3.1224\,(24)$~fm. \label{tab:si}}
\begin{ruledtabular}
\begin{tabular}{lld}
\multicolumn{1}{l}{Effect}
& \multicolumn{1}{l}{Order}   & \multicolumn{1}{c}{$\delta g \times 10^6$} \\
\hline\\[-5pt]
\multicolumn{1}{l}{Electron-electron interaction}
&$1/Z^0$      & -1\,745.249\,323 \\
&$1/Z^1$      &     321.590\,803 \\
&$1/Z^2$      &      -6.876\,0\,(5) \\
&$1/Z^{3+}$   &       0.094\,2\,(11) \\[4pt]
\multicolumn{1}{l}{Finite nuclear size}
&$1/Z^0$      & 0.002\,574\,(4) \\
&$1/Z^1$      & -0.000\,527\,(1)\\[4pt]
\multicolumn{1}{l}{One-loop QED}
&$1/Z^0$      &  1.224\,449\,(3) \\
&$1/Z^1$      & -0.246\,0\,(6) \\
&$1/Z^{2+}$   &  0.009\,6\,(6) \\[4pt]
\multicolumn{1}{l}{Two-loop QED}
&$1/Z^0$      & -0.001\,9\,(3) \\
&$1/Z^1$      &  0.000\,3\,(1) \\[4pt]
\multicolumn{1}{l}{Three-loop QED}
&$1/Z^0$      & 0.000\,013\,(3) \\[4pt]
\multicolumn{1}{l}{Recoil}
&$1/Z^0$      &  0.051\,510\,(1) \\
&$1/Z^{1}$    & -0.007\,585      \\
&$1/Z^{2+}$   & -0.000\,256\,(3) \\[4pt]
\multicolumn{1}{l}{Radiative recoil}
&$1/Z^0$      & -0.000\,040\,(2) \\
&$1/Z^{1+}$   & 0.000\,004 \\[4pt]
\multicolumn{1}{l}{Quadratic recoil}
&$1/Z^0$      & -0.000\,015\,(1) \\[4pt]
\multicolumn{1}{l}{Theory $g-g_e$ (2020) this work}     &&              -1\,429.408\,1\,(15) \\
\multicolumn{1}{l}{Theory $g-g_e$ (2019) \cite{glazov:19} }   &&         -1\,429.410\,0\,(34) \\
\multicolumn{1}{l}{Theory $g-g_e$ (2014) \cite{volotka:14} }  &&         -1\,429.412\,(8) \\
\multicolumn{1}{l}{Experiment \cite{glazov:19} } &&               -1\,429.415\,91\,(14)
%
\end{tabular}
\end{ruledtabular}
\end{table}


\appendix

\begin{widetext}

\section{Perturbation of the one-photon exchange correction}
\label{sec:app1}

In this Section we consider the first-order perturbation of the one-photon exchange correction to
the $g$ factor of the ground state of a Li-like ion by some potential $U$. The result can be used
to derive formulas for the perturbed-orbital self-energy and vacuum-polarization contributions.

The one-photon exchange correction to the $g$ factor of a Li-like ion was derived in
Ref.~\cite{shabaev:02:li} (see also Ref.~\cite{yerokhin:16:gfact:pra}) and can be expressed in
the following form,
\begin{align}
\Delta g_{\rm 1ph} = 2 \sum_{\mu_{c}} \Big[ \Lambda_{\rm 1ph}(vcvc)+\Lambda_{\rm 1ph}(cvcv)
                                                  -\Lambda_{\rm 1ph}(cvvc)-\Lambda_{\rm 1ph}(vccv) \Big]
\,,
\end{align}
where the summation runs over the angular-momentum projection of the core electron states $\mu_c$
and
\begin{align}
\Lambda_{\rm 1ph}(abcd) = \sum_{n\ne a} \frac{\lbr a | V_g | n\rbr \lbr nb| I | cd\rbr}{\vare_a-\vare_n}
+ \frac14\, \lbr ab| I'|cd\rbr\,\Big( \lbr d|V_g|d\rbr - \lbr b|V_g|b\rbr\Big)
\,.
\end{align}
Here, $I \equiv I(\Delta_{db})$ is the operator of the electron-electron interaction and the
prime on $I'$ denotes the derivative over the energy argument.

We now consider the first-order perturbation of $\Lambda_{\rm 1ph}(abcd)$ induced by some
potential $U$. One perturbs the external wave functions, the electron propagator, and the energy
argument of the electron-electron interaction operator $I(\Delta)$ with help of identities
derived in Appendix~\ref{sec:pert}, specifically, Eqs.~(\ref{ap0}), (\ref{ap0a}) and (\ref{ap2}).
The result is conveniently represented as a sum of two parts,
\begin{align}
\delta_U \Lambda_{\rm 1ph}(abcd) = \Lambda_{\rm irred}(abcd) + \Lambda_{\rm red}(abcd)
\,.
\end{align}
The first part $\Lambda_{\rm irred}(abcd)$ contains all terms that do not vanish within the Breit
approximation (i.e., do not contain derivatives of the electron-electron interaction operator),
\begin{align}
\Lambda_{\rm irred}(abcd) &\ =
  \sum_{n_1n_2}{\!}^{'}
  \Bigg\{
  \frac{\lbr a | U | n_1\rbr\, \lbr n_1 | V_g | n_2\rbr\, \lbr n_2 b| I | cd\rbr }{(\vare_a - \vare_{n_1})(\vare_a - \vare_{n_2})}
  + \frac{\lbr a | V_g | n_1\rbr\, \lbr b | U | n_2\rbr\, \lbr n_1n_2 | I | cd\rbr }{(\vare_a - \vare_{n_1})(\vare_b - \vare_{n_2})}
  \nonumber \\ &
  + \frac{\lbr a | V_g | n_1\rbr\,  \lbr n_1b | I | n_2d\rbr\, \lbr n_2 | U | c\rbr\, }{(\vare_a - \vare_{n_1})(\vare_c - \vare_{n_2})}
  + \frac{\lbr a | V_g | n_1\rbr\,  \lbr n_1b | I | cn_2\rbr\, \lbr n_2 | U | d\rbr\, }{(\vare_a - \vare_{n_1})(\vare_d - \vare_{n_2})}
  + \frac{\lbr a | V_g | n_1\rbr\, \lbr n_1 | U | n_2\rbr\, \lbr n_2b | I | cd\rbr }{(\vare_a - \vare_{n_1})(\vare_a - \vare_{n_2})}
  \Bigg\}
  \nonumber \\ &
  - \sum_{n}{\!}^{'}
  \Bigg\{
    \frac{\lbr a | V_g | n\rbr\, \lbr n | U | a\rbr\, \lbr ab | I | cd\rbr\,}{(\vare_a-\vare_{n})^2}
  + \frac{\lbr a | V_g | n\rbr\, \lbr nb | I | cd\rbr\, \lbr a | U | a\rbr\,}{(\vare_a-\vare_{n})^2}
  + \frac{\lbr a | U | n\rbr\, \lbr nb | I | cd\rbr\, \lbr a | V_g | a\rbr\,}{(\vare_a-\vare_{n})^2}
  \Bigg\}
\,.
\end{align}
The second part $\Lambda_{\rm red}(abcd)$ contains terms
with derivatives of the electron-electron interaction,
\begin{align}
\Lambda_{\rm red}(abcd) = &\
   \sum_{n}{\!}^{'}
    \frac{\lbr a | V_g | n\rbr\, \lbr nb | I' | cd\rbr\,}{\vare_a-\vare_{n}}
    \Big( \lbr d | U | d\rbr - \lbr b | U | b\rbr \Big)
   + \sum_{n}{\!}^{'}
    \frac{\lbr a | U | n\rbr\, \lbr nb | I' | cd\rbr\,}{\vare_a-\vare_{n}}
    \Big( \lbr d | V_g | d\rbr - \lbr b | V_g | b\rbr \Big)
    \nonumber \\ &
   + \sum_{n}{\!}^{'}
    \frac{\lbr a | U | n\rbr\,\lbr n | V_g | a\rbr
    }{\vare_a-\vare_{n}}\, \lbr ab | I' | cd\rbr
    + \frac14\, \lbr ab | I'' | cd\rbr
    \Big( \lbr d | V_g | d\rbr - \lbr b | V_g | b\rbr \Big)
    \Big( \lbr d | U | d\rbr - \lbr b | U | b\rbr \Big)
\,.
\end{align}
Note that these formulas are symmetric with respect to $U \leftrightarrow V_g$. So, we could
first perturb $\lbr ab|I|cd\rbr$ with $U$ and then with $V_g$, arriving at the same formulas.

The obtained expressions are equivalent to Eqs.~(51)-(57) and (62)-(64) of Ref.~\cite{glazov:10}
after the substitution $U \to U_{\rm VP}^{\rm el}$, and to Eqs.~(19)-(29) of
Ref.~\cite{glazov:10} after the substitution $U \to \Sigma$. Rewriting the above formulas in an
equivalent way and assuming $U \to \Sigma$, we obtain expression for the perturbed-orbital
self-energy corrections (\ref{eq000})-(\ref{eq0ac}).

\section{Perturbations of energy, wave function, and propagator}
\label{sec:pert}

Let us consider the first-order perturbations of the energy, the wave function, and the electron
propagator induced by a potential $U$. The energy and the wave function obtain the corrections of
the standard form,
\begin{align}\label{ap0}
\delta_U \vare_a &\ = \lbr a|U|a\rbr\,,\\
\delta_U |a\rbr &\ \equiv |\delta_U a \rbr= \sum_{k \ne a} \frac{|k\rbr \lbr k| U|a\rbr}{\vare_a -\vare_k}\,.
\label{ap0a}
\end{align}

We now evaluate the first-order perturbation of the electron propagator with the reference-state
contribution omitted (i.e., the reduced Green function). Perturbing the intermediate-state wave
functions $|k\rbr$ and energies $\vare_a$ and $\vare_k$, we obtain
\begin{align} \label{ap1}
\delta_U \bigg( \sum_{k \neq a} \frac{| k \rbr \lbr k |}{\vare_a - \vare_k} \bigg) =
 \sum_{{k\neq a}\atop{l\neq k}}
 \frac{| k \rbr \lbr k | U | l \rbr \lbr l | }{(\vare_k - \vare_l)(\vare_a - \vare_k)}
 + \sum_{{k\neq a}\atop{l\neq k}}
 \frac{| l \rbr \lbr l | U | k \rbr \lbr k | }{(\vare_k - \vare_l)(\vare_a - \vare_k)}
 + \sum_{k\neq a}\bigl(\lbr k | U | k \rbr - \lbr a | U | a \rbr\bigr)
 \frac{| k \rbr \lbr k | }{(\vare_a - \vare_k)^2}\,.
\end{align}

Rearranging terms, we get
\begin{align} \label{ap1b}
  \sum_{{k\neq a}\atop{l\neq k}}
 \frac{| k \rbr \lbr k | U | l \rbr \lbr l | }{(\vare_k - \vare_l)(\vare_a - \vare_k)}
 + \sum_{{k\neq a}\atop{l\neq k}}
 \frac{| l \rbr \lbr l | U | k \rbr \lbr k | }{(\vare_k - \vare_l)(\vare_a - \vare_k)}
= &\
\sum_{{l,k\neq a}\atop{l\neq k}}
 \frac{| k \rbr \lbr k | U | l \rbr \lbr l | }{(\vare_k - \vare_l)(\vare_a - \vare_k)}
+
\sum_{{l,k\neq a}\atop{l\neq k}}
 \frac{| k \rbr \lbr k | U | l \rbr \lbr l | }{(\vare_l - \vare_k)(\vare_a - \vare_l)}
\nonumber \\ &
- \sum_{k\neq a}
 \frac{| k \rbr \lbr k | U | a \rbr \lbr a | }{(\vare_k - \vare_a)^2}
- \sum_{k\neq a}
 \frac{| a \rbr \lbr a | U | k \rbr \lbr k | }{(\vare_k - \vare_a)^2}
\nonumber \\
= &\ \sum_{{l,k\neq a}\atop{l\neq k}}
 \frac{| k \rbr \lbr k | U | l \rbr \lbr l | }{(\vare_a - \vare_k)(\vare_a - \vare_l)}
- \sum_{k\neq a}
 \frac{| k \rbr \lbr k | U | a \rbr \lbr a | }{(\vare_k - \vare_a)^2}
- \sum_{k\neq a}
 \frac{| a \rbr \lbr a | U | k \rbr \lbr k | }{(\vare_k - \vare_a)^2}\,.
\end{align}

Finally, we obtain
\begin{align} \label{ap2}
\delta_U \bigg( \sum_{k \neq a} \frac{| k \rbr \lbr k |}{\vare_a - \vare_k} \bigg)
= \sum_{l,k\neq a}
 \frac{| k \rbr \lbr k | U | l \rbr \lbr l | }{(\vare_a - \vare_k)(\vare_a - \vare_l)}
- \sum_{k\neq a}
 \frac{| k \rbr \lbr k | U | a \rbr \lbr a | + | a \rbr \lbr a | U | k \rbr \lbr k | }{(\vare_k - \vare_a)^2}
 - \lbr a | U | a \rbr \sum_{k\neq a}
 \frac{| k \rbr \lbr k | }{(\vare_a - \vare_k)^2}\,.
\end{align}

For completeness, we present here also the first-order perturbation of the {\em full} propagator,
\begin{align} \label{ap3}
\delta_U \bigg( \sum_{k} \frac{| k \rbr \lbr k |}{\vare_a - \vare_k} \bigg)
= \sum_{l,k}
 \frac{| k \rbr \lbr k | U | l \rbr \lbr l | }{(\vare_a - \vare_k)(\vare_a - \vare_l)}
 - \lbr a | U | a \rbr \sum_{k}
 \frac{| k \rbr \lbr k | }{(\vare_a - \vare_k)^2}\,.
\end{align}

\section{Free screened vertex contribution}
\label{app:verscr}

The matrix element of the zero-potential two-electron vertex operator can be written in the
momentum space  as (cf. Eqs.~(59) and (60) of Ref.~\cite{yerokhin:99:sescr}),
\begin{align}\label{eq9aa}
\big< ab\big|\Lambda^{(0)}_{\rm scr}\big|cd\big> = \alpha^2 \int \frac{d\bfp\, d\bfp'}{(2\pi)^6}
 \,
 \overline{\psi}_a(\bfp)\, A^{\mu}_{bd}(\Delta_{db},\bfq)\, \Gamma_{R, \mu}(\bfp,\bfp')\,
   \psi_{c}(\bfp')\,,
\end{align}
where $\overline{\psi} = \psi^{\dag}\gamma^0$, $\bfq = \bfp-\bfp'$, $\Delta_{db} =
\vare_d-\vare_b$, $\Gamma_{R, \mu}(\bfp,\bfp')$ is the renormalized free vertex operator (see
Appendix A of Ref.~\cite{yerokhin:99:sescr}) and the 4-vector potential $A^{\mu}$ in the Feynman
gauge is given by
\begin{align}
A^{\mu}_{bd}(\Delta,\bfq) = \frac{4\pi}{\bfq^2-\Delta^2-i0}\,
  \int d\bfz\, \psi_b^{\dag}(\bfz)\, \alpha^{\mu}\, \psi_d(\bfz)\, e^{-i\bfq\cdot\bfz}\,.
\end{align}

Performing the angular integration over $\hat{\bm{z}}$ for $A^{\mu} = (A^0,{\bf A})$, we obtain
(see Eqs.(143) and (144) of Ref.~\cite{yerokhin:03:epjd})
\begin{eqnarray}  \label{OPf3}
A_{bd}^0(\Delta,\bfq) &=& \frac{16\pi^2}{\bfq^2-\Delta^2-i0}
      \sum_{JM} i^{-J}
      s_{JM}^{db}\, Y_{JM}(\hq)\,
         P^{1,bd}_J(q)\,,
\end{eqnarray}
\begin{eqnarray} \label{OPf4}
{\bf A}_{bd}(\Delta,\bfq) &=& \frac{16\pi^2}{\bfq^2-\Delta^2-i0}
      \sum_{JLM} i^{1-L}
      s_{JM}^{db}\,
      {\bf Y}_{JLM}(\hq)\,   P^{2,bd}_{JL}(q)\,,
\end{eqnarray}
where $\hq \equiv \bfq/|\bfq|$,
\begin{align}
 s_{LM}^{db} = \frac{(-1)^{j_b-\mu_b}}{\sqrt{4\pi}}\, C^{LM}_{j_d\mu_d, j_b -\mu_b}
             \, ,
\end{align}
$Y_{Lm}$ are the spherical harmonics,  ${\bf Y}_{JLM}$ are the vector spherical harmonics,
\begin{align}
{\bf Y}_{JLM}(\hat{\bfx}) = \sum_{m q} C^{JM}_{Lm,1q}\, Y_{Lm}(\hat{\bfx})\,
        {\bf e}_q \,,
\end{align}
and ${\bf e}_q$ are the spherical components of the unity vector. The radial integrals are
defined as
\begin{align}
P^{1,bd}_J(q) =&\ C_J(\kappa_d,\kappa_b) \int_0^{\infty} dx\, x^2 j_J(q x)\big(g_bg_d+f_bf_d\big)\,,
    \\
P^{2,bd}_{JL}(q) =&\  \int_0^{\infty} dx\, x^2 j_L(q x)
 \Bigl[g_bf_d  S_{JL}(\kappa_b,-\kappa_d)
    -f_bg_d S_{JL}(-\kappa_b,\kappa_d)\Bigr]\,,
\end{align}
where  $g_i=g_i(x)$ and $f_i = f_i(x)$ are components of radial wave functions, $j_l(z)$ is the
spherical Bessel function, and the angular coefficients $C_{J}(\kappa_1,\kappa_2)$ and
$S_{JL}(\kappa_1,\kappa_2)$ are given by Eqs.~(274)-(277) of Ref.~\cite{yerokhin:03:epjd}.

In order to perform angular integrations over $\hat{\bfp}_1$ and $\hat{\bfp}_2$ in
Eq.~(\ref{eq9aa}), we use the following representation for the vertex operator sandwiched between
two Dirac wave functions \cite{yerokhin:99:sescr}
\begin{align}
 \overline{\psi}_a(\bfp_1)\,
\Gamma_{R,0}(\bfp_1,\bfp_2)\,
   \psi_c(\bfp_2)
      =   \frac{\alpha}{4\pi}\, i^{l_a-l_c}\, \Bigl[
        {\cal F}^{ac}_1 \chi^{\dag}_{\kappa_a \mu_a}(\hp_1)
        \chi_{\kappa_c \mu_c}(\hp_2)
 +        {\cal F}^{ac}_2 \chi^{\dag}_{-\kappa_a \mu_a}(\hp_1)
        \chi_{-\kappa_c \mu_c}(\hp_2) \Bigr] \, ,
\end{align}
\begin{align}
        \label{vertex3a}
 \overline{\psi}_a(\bfp_1)\,
{\mbox{\boldmath$\Gamma$}}_R(\bfp_1,\bfp_2)\,
        \psi_c(\bfp_2)
       =  &\
       \frac{\alpha}{4\pi} \, i^{l_a-l_c} \Bigl[
    {\cal R}^{ac}_1 \chi^{\dag}_{\kappa_a \mu_a}(\hat{\bfp}_1)
        {\bsigma} \chi_{-\kappa_c\mu_c}(\hat{\bfp}_2)
   +{\cal R}^{ac}_2 \chi^{\dag}_{-\kappa_a \mu_a}(\hat{\bfp}_1)
        {\bsigma} \chi_{\kappa_c\mu_c}(\hat{\bfp}_2)
         \nonumber \\
&+ \left({\cal R}^{ac}_3 \bfp_1+ {\cal R}^{ac}_4 \bfp_2\right)
         \chi^{\dag}_{\kappa_a \mu_a}(\hat{\bfp}_1)
         \chi_{\kappa_c\mu_c}(\hat{\bfp}_2)
   +  \left({\cal R}^{ac}_5 \bfp_1 +{\cal R}^{ac}_6
\bfp_2\right)
         \chi^{\dag}_{-\kappa_a \mu_a}(\hat{\bfp}_1)
         \chi_{-\kappa_c\mu_c}(\hat{\bfp}_2) \Bigr] \,,
\end{align}
where $\chi_{\kappa\mu}(\hat{\bfp})$ are the spin-angular spinors. The one-loop vertex functions
${\cal F}^{ac}_i = {\cal F}^{ac}_i(p_{1},p_{2},\xi)$ and ${\cal R}^{ac}_i = {\cal
R}^{ac}_i(p_{1},p_{2},\xi)$ are given in Appendix A of Ref.~\cite{yerokhin:99:sescr}, with $q=
|\bfq|$, $p_{1} = |\bfp_1|$, $p_{2} = |\bfp_2|$, and $\xi = \hp_1\cdot \hp_2$.

Now we perform integrations over all angular variables except for $\xi$ in Eq.~(\ref{eq9aa}), by
using the following identity:
\begin{align}
\int d\hp_1\,d\hp_2\, F(p_1,p_2,\xi)\, G(\hp_1,\hp_2) = \int_{-1}^1 d\xi\,F(p_1,p_2,\xi)\,g(\xi)\,,
\end{align}
where $F(p_1,p_2,\xi)$ and $G(\hp_1,\hp_2)$ are arbitrary functions of the specified arguments,
and
\begin{align}
g(\xi) = 2\pi \sum_{lm} P_l(\xi)\,  \int d\hp_1\,d\hp_2\, Y_{lm}(\hp_1)\, Y^*_{lm}(\hp_2)\,G(\hp_1,\hp_2)\,,
\end{align}
where $P_l$  are the Legendre polynomials. Using this identity, the matrix element of the free
two-electron vertex can be expressed \cite{yerokhin:99:sescr} (see also Sec.~3.2 of
Ref.~\cite{yerokhin:03:epjd}) as
\begin{align}
\lbr ab|\Lambda^{(0)}_{\rm scr}|cd\rbr
 = &\  \frac{\alpha^2}{2\pi^3}\,
 \int_0^{\infty} dp_1\,dp_2\,\int_{-1}^1 d\xi\,
   \frac{p_1^2p_2^2}{\bfq^2-\Delta_{db}^2-i0}\,
\nonumber \\ & \times
\sum_J
    \Bigg\{ i^{J-l_a+l_c}\,
      P^{1}_J(q,bd)\, \Big[ {\cal F}^{ac}_1\,t_{l_al_c}(J)
                            + {\cal F}^{ac}_2\, t_{\overline{l}_a\overline{l}_c}(J) \Big]
\nonumber \\ &
 - \sum_L i^{L-l_a+l_c-1}\, P^{2}_{JL}(q,bd)\,
       \Big[ {\cal R}^{ac}_1\,s_{l_a\overline{l}_c}^{\sigma}(JL)
             + {\cal R}^{ac}_2\,s_{\overline{l}_al_c}^{\sigma}(JL)
\nonumber \\ &
+ p_{1}\, {\cal R}^{ac}_3\,s_{l_al_c}^{p_1}(JL)
+ p_{2}\, {\cal R}^{ac}_4\,s_{l_al_c}^{p_2}(JL)
+ p_{1}\, {\cal R}^{ac}_5\,s_{\overline{l}_a\overline{l}_c}^{p_1}(JL)
+ p_{2}\, {\cal R}^{ac}_6\,s_{\overline{l}_a\overline{l}_c}^{p_2}(JL)
\Big]
\Bigg\}
\,,
\end{align}
where $l_i = |\kappa_i+1/2|-1/2$, $\overline{l}_i = |\kappa_i-1/2|-1/2$.   The angular factors
$t_{l_1l_2}$, $s_{l_1l_2}^{\sigma}$ and $s_{l_1l_2}^{p_i}$ are defined as follows:
\begin{align}\label{eq:t}
t_{l_al_c}(J) =&\ \frac{1}{4\pi}\,\sum_{lm} P_l(\xi)\, \int d\hp_1\,d\hp_2\,
  Y_{lm}(\hp_1)\,Y_{lm}^*(\hp_2)\,
   \sum_M\, s^{db}_{JM}\, \chi^{\dag}_{\kappa_a\mu_a}(\hp_2)\,Y_{JM}(\hq)\,
                         \chi_{\kappa_c\mu_c}(\hp_1)\,,
\end{align}
\begin{align}\label{eq:ssi}
s_{l_al_c}^{\sigma}(JL) =&\ \frac{1}{4\pi}\,\sum_{lm} P_l(\xi)\, \int d\hp_1\,d\hp_2\,
  Y_{lm}(\hp_1)\,Y_{lm}^*(\hp_2)\,
   \sum_{M}\, s^{db}_{JM}\,
     \,\chi^{\dag}_{\kappa_a\mu_a}(\hp_2)\,{\bm{\sigma}}\cdot {\bm Y}_{JLM}(\hq)
                         \chi_{\kappa_c\mu_c}(\hp_1)\,,
\end{align}
\begin{align}\label{eq:sp}
s_{l_al_c}^{p_i}(JL) =&\ \frac{1}{4\pi}\,\sum_{lm} P_l(\xi)\,  \int d\hp_1\,d\hp_2\,
  Y_{lm}(\hp_1)\,Y_{lm}^*(\hp_2)\,
   \sum_{M}\, s^{db}_{JM}\,
     \,\chi^{\dag}_{\kappa_a\mu_a}(\hp_2)\,\hp_i\cdot {\bm Y}_{JLM}(\hq)
                         \chi_{\kappa_c\mu_c}(\hp_1)\,.
\end{align}

The angular coefficients (\ref{eq:t})-(\ref{eq:sp}) appeared previously in the calculation of the
two-loop self-energy (see Eqs.~(150) and (151) of Ref.~\cite{yerokhin:03:epjd}), but there they
were averaged over the angular-momentum projections of the reference states, which allowed to
simplify expressions considerably. In the present work we evaluate these angular coefficients in
the general case. Using the standard Racah angular momentum algebra, we obtain the following
results:
\begin{align}
t_{l_al_c}(J) =&\ \frac{1}{4\pi}\,\sum_{ll_1} P_l(\xi)\, c_{l_1l_2}^J
\sum_{m\sigma}
    s_{JM}^{db}\, C_{l_1m_1,l_2m_2}^{JM}\,
                  C_{l_am_a,1/2\,\sigma}^{j_a\mu_a}\,
                  C_{l_cm_c,1/2\,\sigma}^{j_c\mu_c}\,
                  R_3(l_a,m_a,l,m,l_1,m_1)\,
                  R_3(l,m,l_c,m_c,l_2,m_2)\,
                  ,
\end{align}
\begin{align}
s_{l_al_c}^{\,\sigma}(JL) =&\ \frac{1}{4\pi}\,\sum_{ll_1} P_l(\xi)\,  c_{l_1l_2}^L
\sum_{m\sigma_a\sigma_c}
    s_{JM}^{db}\, C_{Lm_L,1q}^{JM}\,
                  C_{l_am_a,1/2\,\sigma_a}^{j_a\mu_a}\,
                  C_{l_cm_c,1/2\,\sigma_c}^{j_c\mu_c}\,
\nonumber \\ & \times
    \sqrt{2}\,(-1)^{1/2-\sigma_c}\,C^{1q}_{1/2-\sigma_c,1/2\,\sigma_a}\,
    C_{l_1m_1,l_2m_2}^{Lm_L}\,
                  R_3(l_a,m_a,l,m,l_1,m_1)\,
                  R_3(l,m,l_c,m_c,l_2,m_2)\,
                  ,
\end{align}
\begin{align}
s_{l_al_c}^{p_1}(JL) =&\ \frac{1}{\sqrt{12\pi}}\,\sum_{ll_1} P_l(\xi)\,  c_{l_1l_2}^L
\sum_{m\sigma q}
    s_{JM}^{db}\, C_{Lm_L,1q}^{JM}\,
                  C_{l_am_a,1/2\,\sigma_a}^{j_a\mu_a}\,
                  C_{l_cm_c,1/2\,\sigma_c}^{j_c\mu_c}\,
\nonumber \\ & \times
    C_{l_1m_1,l_2m_2}^{Lm_L}\,
                  R_4(l_a,m_a,l,m,l_1,m_1,1,q)\,
                  R_3(l,m,l_c,m_c,l_2,m_2)\,
                  ,
\end{align}
where
\begin{align}
 R_3(l_1,m_1,l_2,m_2,l_3,m_3) = \int d\hx\, Y^*_{l_1m_1}(\hx)\,Y_{l_2m_2}(\hx)\,Y_{l_3m_3}(\hx)\,,
\end{align}
\begin{align}
 R_4(l_1,m_1,l_2,m_2,l_3,m_3,l_4,m_4) = \int d\hx\, Y^*_{l_1m_1}(\hx)\,Y_{l_2m_2}(\hx)\,Y_{l_3m_3}(\hx)\,Y_{l_4m_4}(\hx)\,,
\end{align}
and $c_{l_1l_2}^L$ are coefficients of the expansion of the spherical harmonics of
$\hat{\bfz}\equiv \hat{\bfz}_1-\hat{\bfz}_2$ into spherical harmonics of $\hat{\bfz}_1$ and
$\hat{\bfz}_2$  \cite{varshalovich}
\begin{equation} \label{vector2}
Y_{LM}(\hat{\bfz}) =  \sum_{{l_1,l_2=0}\atop{l_1+l_2=L}}^{L}
        c^L_{l_1l_2}
        \sum_{m_1m_2}
        C^{LM}_{l_1m_1,l_2m_2} Y_{l_1m_1}(\hat{\bfz}_1)
        Y_{l_2m_2}(\hat{\bfz}_2)\ ,
\end{equation}
with
\begin{equation} 
c^L_{l_1l_2} =
        \sqrt{\frac{4\pi(2L+1)!}{(2l_1+1)!(2l_2+1)!}}
        \frac{z_1^{l_1}z_2^{l_2}}{z^L} \,(-1)^{l_2} \ .
\end{equation}
The integrals of three and four spherical harmonics $R_3$ and $R_4$ are evaluated in terms of
Clebsch-Gordan coefficients by standard formulas \cite{varshalovich}.

For specific cases, the angular coefficients can be evaluated analytically, as it was done in
Ref.~\cite{yerokhin:99:sescr}. In the present work we prefer to evaluate all sums of
Clebsch-Gordan coefficients numerically. The specific cases of $J = 0$ and $J = 1$ were
simplified and evaluated separately.

We now consider the matrix element of the zero-potential two-electron vertex operator for the
Coulomb gauge of the exchanged photon. In this case the 4-vector potential $A^{\mu} = (A^0,A^i)$
becomes
\begin{align}
A^{0}_{bd}(\Delta,\bfq) &\, = \frac{4\pi}{\bfq^2-i0}\,
  \int d\bfz\, \psi_b^{\dag}(\bfz)\, \psi_d(\bfz)\, e^{-i\bfq\cdot\bfz}\,,
  \nonumber \\
A^{i}_{bd}(\Delta,\bfq) &\, = \Big( \delta_{ij} - \frac{q_i\,q_j}{q^2}\Big)\,
    \frac{4\pi}{\bfq^2-\Delta^2- i0}\,
  \int d\bfz\, \psi_b^{\dag}(\bfz)\, \alpha_j\,\psi_d(\bfz)\, e^{-i\bfq\cdot\bfz}\,.
\end{align}
In order to perform the angular integrations in the momentum integrations, we use the following
representation
\begin{align}
 \overline{\psi}_a(\bfp_1)\,
\frac{
\bfq\, \big( \bfq\cdot \bm{\Gamma}_R \big)}{q^2}\,
        \psi_c(\bfp_2)
=
       \frac{\alpha}{4\pi} \, i^{l_a-l_c} \Bigl[
\widetilde{{\cal R}}^{ac}_3\,\bfq\,
         \chi^{\dag}_{\kappa_a \mu_a}(\hat{\bfp}_1)
         \chi_{\kappa_c\mu_c}(\hat{\bfp}_2)
   +  \widetilde{{\cal R}}^{ac}_5 \, \bfq\,
         \chi^{\dag}_{-\kappa_a \mu_a}(\hat{\bfp}_1)
         \chi_{-\kappa_c\mu_c}(\hat{\bfp}_2) \Bigr] \,,
\end{align}
where the functions $\widetilde{{\cal R}}_i$ are expressed in terms of functions ${\cal R}_i$ as
follows:
\begin{align}
\widetilde{{\cal R}}^{ac}_3 &\, = \frac1{q^2}
\Big[ p_2\, {\cal R}^{ac}_1 - p_1\,{\cal R}^{ac}_2 + {\cal R}^{ac}_3\big(p_1^2-p_{12}\big)
   + {\cal R}^{ac}_4\big(p_{12}-p_2^2\big)\Big]\,,
   \\
\widetilde{{\cal R}}^{ac}_5 &\, = \frac1{q^2}
\Big[ -p_1\, {\cal R}^{ac}_1 + p_2\,{\cal R}^{ac}_2 + {\cal R}^{ac}_5\big(p_1^2-p_{12}\big)
   + {\cal R}^{ac}_6\big(p_{12}-p_2^2\big)\Big]\,,
\end{align}
where $p_{12} = \bfp_1 \cdot \bfp_2$.  We observe that the matrix element in the Coulomb gauge
involves the same angular coefficients (\ref{eq:t})-(\ref{eq:sp}) as in the Feynman gauge.
\end{widetext}

\end{document}